\documentclass[pdflatex,sn-nature]{sn-jnl}

\usepackage{graphicx}%
\usepackage{multirow}%
\usepackage{amsmath,amssymb,amsfonts}%
\usepackage{amsthm}%
\usepackage{mathrsfs}%
\usepackage[title]{appendix}%
\usepackage{xcolor}%
\usepackage{textcomp}%
\usepackage{manyfoot}%
\usepackage{booktabs}%
\usepackage{algorithm}%
\usepackage{algorithmicx}%
\usepackage{algpseudocode}%
\usepackage{listings}%
\usepackage{colortbl}

\usepackage{subfigure}
\usepackage[T1]{fontenc}

\usepackage{color, colortbl}
\definecolor{Gray}{gray}{0.9}
\usepackage{amsmath}
\usepackage{pifont}

\usepackage{makecell}
\usepackage{wrapfig}
\usepackage{picinpar}

\usepackage[utf8]{inputenc} %
\usepackage{hyperref}       %
\usepackage{url}            
\usepackage{amsfonts}       %
\usepackage{nicefrac}       %
\usepackage{microtype}      %

\usepackage{bm}

\usepackage{CJKutf8}

\usepackage{colortbl,xcolor}
\usepackage[most]{tcolorbox}
\newtcolorbox[list inside=prompt]{prompt}[1][]{
    enhanced,
    colback=green!5!gray!10,
    colframe=green!20!gray!55,
    colbacktitle=green!20!gray!55,
    coltitle=white,
    fonttitle=\bfseries\sffamily,
    fontupper=\small\ttfamily,
    title=Prompt,
    attach boxed title to top center={yshift=-3pt},
    boxed title style={
        rounded corners=northeast,
        rounded corners=southeast,
        boxrule=0pt
    },
    rounded corners,
    boxrule=1.5pt,
    left=2pt,          
    right=0pt,         
    top=5pt,           
    bottom=0pt,        
    boxsep=2pt,        
    before skip=5pt,   
    after skip=5pt,    
    breakable,
    #1,
}
\definecolor{greyblue}{RGB}{120,145,180}
\newtcolorbox[list inside=prompt]{encode}[1][]{
    enhanced,
    colback=greyblue!15,
    colframe=greyblue!60,
    colbacktitle=greyblue!60,
    coltitle=white,
    fonttitle=\bfseries\sffamily,
    fontupper=\small\ttfamily,
    title=Prompt,
    attach boxed title to top center={yshift=-3pt},
    boxed title style={
        rounded corners=northeast,
        rounded corners=southeast,
        boxrule=0pt
    },
    rounded corners,
    boxrule=1.5pt,
    left=2pt,          
    right=0pt,         
    top=5pt,           
    bottom=0pt,        
    boxsep=2pt,        
    before skip=5pt,   
    after skip=5pt,    
    breakable,
    #1,
}
\theoremstyle{thmstyleone}%
\theoremstyle{thmstyletwo}%

\theoremstyle{thmstylethree}%

\begin{document}

\title[Article Title]{LEXA: Legal Case Retrieval via Graph Contrastive Learning with Contextualised LLM Embeddings}


\author{\fnm{Yanran} \sur{Tang}}\email{\{yanran.tang, r.qiu, yilun.liu, helen.huang\}@uq.edu.au}

\author{\fnm{Ruihong} \sur{Qiu}}

\author{\fnm{Yilun} \sur{Liu}}

\author{\fnm{Xue} \sur{Li}}\email{xueli@eecs.uq.edu.au}

\author{\fnm{Zi} \sur{Huang}}

\affil{\orgname{The University of Queensland}, \orgaddress{\country{Australia}}}


\abstract{Legal case retrieval (LCR) is a specialised information retrieval task aimed at identifying relevant cases given a query case. LCR holds pivotal significance in facilitating legal practitioners to locate legal precedents. Existing LCR methods predominantly rely on traditional lexical models or language models; however, they typically overlook the domain-specific structural information embedded in legal documents. Our previous work CaseGNN~\cite{casegnn} successfully harnesses text-attributed graphs and graph neural networks to incorporate structural legal information. Nonetheless, three key challenges remain in enhancing the representational capacity of CaseGNN: (1) The under-utilisation of rich edge information in text-attributed case graph (TACG). (2) The insufficiency of training signals for graph contrastive learning. (3) The lack of contextualised legal information in node and edge features. In this paper, the LEXA model, an extension of CaseGNN, is proposed to overcome these limitations by jointly leveraging rich edge information, enhanced training signals, and contextualised embeddings derived from large language models (LLMs). Specifically, an edge-updated graph attention layer (EUGAT) is proposed to comprehensively update node and edge features during graph modelling, resulting in a full utilisation of structural information of legal cases. Moreover, LEXA incorporates a novel graph contrastive learning objective with graph augmentation to provide additional training signals, thereby strengthening the model's legal comprehension capabilities. What's more, given the remarkable contextualised understanding capabilities of LLMs for text encoding, LLMs are employed to generate node and edge features for the text-attributed case graph (TACG). Extensive experiments on two benchmark datasets from COLIEE 2022 and COLIEE 2023 demonstrate that LEXA not only significantly improves CaseGNN but also achieves supreme performance compared to state-of-the-art LCR methods. Code has been released on \href{https://github.com/yanran-tang/CaseGNN}{here}.
}




\maketitle

\section{Introduction}\label{sec1}
Given a query case, legal case retrieval (LCR) aims to find relevant cases in a large case database, and has become an indispensable tool for legal practitioners such as judges and lawyers. Recently, the development of LCR tools has significantly streamlined this process, enabling rapid and accurate retrieval of relevant cases. Moreover, LCR tools also serve as a convenient resource for individuals seeking legal consultation but lacking the financial means to afford expensive legal services. In recent years, LCR approaches have been mostly based on two methodologies: lexical retrieval models and language models (LMs). Traditional lexical retrieval models, including BM25~\cite{BM25}, TF-IDF~\cite{TF-IDF} and LMIR~\cite{LMIR}, calculate the similarity score between cases based on term frequency. Language models~\cite{Law2Vec,Lawformer,MTFT-BERT,MVCL,BERT-PLI,LEGAL-BERT,SAILER,JOTR,DoSSIER,RPRS, NOWJ,ConversationalAgent,UA@COLIEE2022,CL4LJP} rely on pre-trained models to encode cases into case representations for calculating case similarity. Specifically, due to the intricate structure of legal texts, language model-based LCR methods employ various case similarity comparison strategies, including sentence-level~\cite{IOT-Match}, paragraph-level~\cite{BERT-PLI}, and document-level~\cite{SAILER} approaches.

However, important legal structural information is neglected in LM-based LCR models, as they rely solely on raw text to generate high-dimensional representations of cases for similarity calculation. The structural information of a legal case captures the relationships among its core elements, such as parties, criminal activities, and evidence, which can help to further improve the understanding capability of the model. Therefore, in our previous work, CaseGNN~\cite{casegnn} is proposed to leverage legal structural relationships to improve LCR accuracy. Firstly, the structural information of legal cases is extracted in the form of relational triplets, which are then encoded by LMs to generate node features for the constructed Text-Attributed Case Graph (TACG). With the constructed TACG, a CaseGNN framework is proposed to generate the final case representation. Finally, to train the CaseGNN model, a contrastive learning loss is designed to capture the training signal from both easy and hard negative samples.

Although CaseGNN exploits the legal structural information that is neglected in the lexical models and LMs, three key challenges remain unaddressed, hindering the accuracy and effectiveness of LCR. \textbf{(1) Underutilisation of rich edge information}: CaseGNN only focuses on updating node features during graph modelling, thereby neglecting the rich and informative edge features inherent in the TACG. The edges in the TACG represent relationships between nodes, capturing legal connections between entities, which are crucial for accurately generating case representations via graph neural network models. However, in the GNN layer of CaseGNN, the edge features remain fixed throughout the training process, preventing the model from leveraging updated relational information derived from evolving node features. \textbf{(2) Insufficiency of training signals}: The training process of the CaseGNN model is hindered by a scarcity of training signals, primarily due to the limited availability of labelled legal data. This challenge is further exacerbated by the restricted access to legal sources and the high cost associated with annotating legal datasets. Since the labelling process typically requires legal expertise from professionals such as judges, lawyers, or legal scholars, it incurs substantial time and financial costs. Consequently, the inadequacy of labelled data leads to insufficient training signals for effectively training LCR models. \textbf{(3) The lack of contextualised legal information in node and edge features}: In CaseGNN, the node and edge features are encoded by LMs, which lacks contextualised legal information. This limitation restricts the model’s ability to capture the intricate legal semantics and interrelations embedded in legal texts, thereby affecting the expressiveness of the final case representations generated from case graphs constructed with suboptimal node and edge features.

To address the identified limitations of CaseGNN, an enhanced framework, LEXA, is proposed to extend CaseGNN~\cite{casegnn} in a more accurate and effective manner. First, LEXA collaboratively exploits and dynamically updates both node and edge features by proposing a novel Edge-updated Graph Attention Layer to generate more comprehensive and informative case representations. Second, to mitigate the issue of insufficient training signals, graph contrastive learning with graph augmentation is employed, providing additional supervision to enhance the model’s representational capacity. Finally, a top-performing text encoding-oriented large language model (LLM) is leveraged to encode node and edge features, capturing contextualised legal semantics embedded within legal cases. Empirical experiments are conducted on two benchmark datasets, COLIEE 2022~\cite{COLIEE2022} and COLIEE 2023~\cite{COLIEE2023}, demonstrating that LEXA not only outperforms the original CaseGNN model but also achieves state-of-the-art performance on the legal case retrieval (LCR) task. Compared with CaseGNN~\cite{casegnn}, the main contributions of this paper are summarised as follows:
\begin{itemize}
    \item A LEXA framework is proposed to address the limitations of CaseGNN and to further improve retrieval accuracy and effectiveness by fully exploiting edge information, providing additional training signals, and leveraging LLMs to incorporate contextualised legal semantics.
    \item A GNN layer called Edge-updated Graph Attention Layer (EUGAT) is designed to simultaneously update both node and edge features. This mechanism enables the generation of comprehensive and informative case representations for model training.
    \item A graph contrastive learning (GCL) objective is developed, along with a graph augmentation strategy to provide supplementary training signals that enhance the model’s legal understanding capabilities.
    \item A top-performing text-embedding LLM, LEXA-8B, is fine-tuned and employed to encode node and edge features, enabling richer contextualised and latent legal semantics to be captured within legal cases.
    \item Extensive experiments conducted on the COLIEE 2022 and 2023 benchmark datasets validate the superior performance of LEXA, demonstrating significant improvements over CaseGNN and outperforming existing state-of-the-art LCR approaches.
\end{itemize}

\section{Related Work}
In this section, the related work will be reviewed in detail from three aspects that related to this paper, which are legal case retrieval, graph contrastive learning, and text embedding LLM.

\subsection{Legal Case Retrieval}
Legal Case Retrieval, as a specialised subtask of information retrieval, has seen the development of two main categories of methods: traditional statistical retrieval models and neural language models. Statistical models such as TF-IDF~\cite{TF-IDF}, BM25~\cite{BM25}, and LMIR~\cite{LMIR} rely on term frequency and inverse document frequency to compute relevance scores between a query and candidate cases. These approaches are efficient and interpretable but often fall short in capturing semantic similarities in complex legal texts. More recently, language models~\cite{DeepCT,BERT,RoBERTa,monot5} have gained prominence in LCR due to their strong language understanding capabilities. Numerous LCR-specific adaptations of LMs have been proposed~\cite{Law2Vec,Lawformer,MTFT-BERT,MVCL,LEGAL-BERT,JOTR,DoSSIER,RPRS,NOWJ,UA@COLIEE2022,IOT-Match,Law-Match,LEDsummary,BM25injtct,LeiBi,LEVEN,JNLP@COLIEE2019,CL4LJP,QAjudge,query_conversational_agent, ConversationalAgent}, demonstrating substantial improvements over traditional retrieval techniques. Specifically, Law2Vec~\cite{Law2Vec} utilises a large number of English legal corpus to train a Word2Vec~\cite{word2vec} model to get legal word embeddings. Besides, Lawformer~\cite{Lawformer} leverages Chinese legal documents to pre-train a longformer~\cite{longformer} based model into a Chinese specific legal language model. MVCL~\cite{MVCL} exploits a contrastive learning model with case-view and element-view strategies to enhance the understanding ability of the model and improve the performance of relevant case retrieval task. And CL4LJP~\cite{CL4LJP} combines three different levels contrastive learning objectives that includes article-level, charge-level and label-aware level to jointly train the legal judgment prediction model. To tackle the long text problem in legal domain, BERT-PLI~\cite{BERT-PLI} divides cases into paragraphs and calculates the similarity between two paragraphs while SAILER~\cite{SAILER} directly truncates the case text to cope with the input limit of LM. PromptCase~\cite{promptcase} proposes a prompt-based encoding scheme with two important legal facts and legal issues to address the lengthy problem. CaseLink~\cite{caselink} utilises the intrinsic case connectivity relationships among cases with graph neural networks to improve the retrieval accuracy.

\subsection{Graph Contrastive Learning}
Contrastive learning, a self-supervised learning method, focuses on learning the feature representation by bringing the positive samples closer while pushing the negative samples away~\cite{nce,NCE4NPLMs}. Contrastive learning is widely used in computer vision task~\cite{moco,SimCLR}, natural language processing task~\cite{SimCSE,CLEAR,COCO-LM}, and information retrieval task~\cite{PSSL,ANCE} for providing additional training signals to train the model instead of extra annotation. 

Recently, contrastive learning is also applied in graph domain to fully leverage the graph structural information and explore the unlabelled graph data. The main difference between contrastive learning in the graph domain and other domains is the augmentation methods of graphs focusing on the aspects of graph structure such as node, edge or subgraph, where other domains have no graph structure. Specifically, there are various graph augmentation methods used for graph contrastive learning. Edge dropping is firstly proposed by Rong et al~\cite{DropEdge}to remove a certain portion of edges from the original graph. Similarly, node dropping~\cite{GRAND,GraphCL} refers to removing nodes with a certain probability. Besides of the graph structure augmentation such as removing nodes and edges, graph feature masking~\cite{BGRL,GraphCL} is also a popular graph augmentation method to perform efficient graph augmenting or corrupting by randomly mask the features of nodes or edges. Considering graph can be divided into subgraph, another graph augmentation approach called subgraph cropping is to crop out subgraph from the original graph then use the subgraph itself or the remaining graph as samples for contrastive learning (e.g., ~\cite{GraphCL, GraphCrop, puma, tnt-ood}).

\subsection{LLM Embeddings}
Large language models (LLMs) have demonstrated outstanding performance in encoding text into high-quality embeddings for a wide range of natural language processing tasks, such as information retrieval and question answering. Their impressive encoding capability arises from the ability to model complex linguistic patterns and contextual dependencies through deep transformer architectures pretrained on large-scale corpora. Various LLM-based embedding models have been proposed to enhance semantic representation across diverse tasks. The recently proposed Multilingual Massive Text Embedding Benchmark (MMTEB)~\cite{mmteb} offers a comprehensive evaluation framework for assessing text embedding models across multiple tasks, such as classification, retrieval, and clustering. Legal retrieval is a domain-specific task included in MMTEB, where several state-of-the-art text embedding LLMs have achieved notable performance. voyage-law-2\footnote{https://blog.voyageai.com/2024/04/15/domain-specific-embeddings-and-retrieval-legal-edition-voyage-law-2/} ranks first in the MMTEB legal retrieval task by leveraging high-quality legal tokens, and a novel contrastive learning algorithm. E5-Mistral-7B-Instruct\cite{e5mistral} is another top-performing model that utilizes proprietary LLMs to generate diverse synthetic data for training. OpenAI’s text-embedding-3-large\footnote{https://platform.openai.com/docs/guides/embeddings} is a high-quality general purpose embedding model, though it is primarily optimized for English and not specifically designed for multilingual retrieval tasks such as those in MMTEB. Additionally, Qwen3-Embedding-8B\cite{qwen3embedding}, trained using an innovative multi-stage pipeline that combines large-scale unsupervised pretraining with supervised fine-tuning on high quality datasets, demonstrates strong performance across both general purpose and domain-specific retrieval tasks. Specifically, some LLM-embedding–based legal retrieval methods, such as UQLegalAI@COLIEE2025~\cite{uqlegalai}, and legally fine-tuned large language models, such as ReaKase-8B~\cite{reakase}, have been proposed to enhance semantic understanding of legal texts. These advancements highlight the growing effectiveness and specialization of LLM-based embedding models in both general and domain-specific retrieval scenarios, particularly within the legal domain.

\section{Preliminary}
In the following, a bold lowercase letter denotes a vector, a bold uppercase letter denotes a matrix, a lowercase letter denotes a scalar or a sequence of words, and a scripted uppercase letter denotes a set.

\subsection{Task Definition}
\label{sec: task}
In legal case retrieval, given the query case $q$, and the set of $n$ candidate cases $\mathcal{D}=\{d_1,d_2,...,d_n\}$, our task is to retrieve a set of relevant cases $\mathcal{D}^* = \{d^*_i| d^*_i \in D \wedge relevant (d^*_i, q) \}$, where $relevant (d^*_i, q)$ denotes that $d^*_i$ is a relevant case of the query case $q$. The relevant cases are called precedents in legal domain, which refer to the historical cases that can support the judgement of the query case.

\subsection{Case Graph Construction}
In CaseGNN, legal cases are transformed into case graphs to capture their underlying structural information. Specifically, legal relations among entities within a case are extracted using information extraction tools, with entities represented as nodes and the extracted relations as edges in the resulting case graph.

\subsubsection{Extraction of Legal Relation and entities.}
\label{sec:extract}
To construct case graphs that reflect the structural characteristics of legal cases, the named entity recognition tools and relation extraction tools to extract the legal relations and entities of cases. For instance, in the COLIEE datasets compiled from the Federal Court of Canada~\cite{COLIEE2022,COLIEE2023}, a sample triplet example ($applicant$, $is$, $Canadian$) is extracted from the sentence ``The applicant is a Canadian'', where $applicant$ corresponds to the ``who'' and $is, Canadian$ to the ``what'' component in the case. Finally, a set of relation triplets $R\ = \{(h,\ r,\ t)_{i=1:n}\}$ is extracted, where $h$ and $t$ denote the head and tail entities, respectively, $r$ represents the relation between them, and $n$ is the total number of triplets identified in a case.

\subsubsection{Case Graph Construction.}
In legal domain, the relevance of cases is primarily determined by the alignment of legal facts and legal issues\cite{promptcase}. Legal facts describe fundamental aspects of a case such as who, when, what, where, and why. And legal issues pertain to the legal disputes between parties that require judicial resolution\cite{promptcase}. Therefore, the case graph of every case is constructed into a legal fact graph and a legal issue graph separately. Firstly, the legal fact and legal issue are extracted by PromptCase module, which is detailed in the PromptCase paper~\cite{promptcase}. Given a legal case $c$, let $R_{c,\text{fact}}$ and $R_{c,\text{issue}}$ denote the sets of relation triplets corresponding to the \textit{legal fact} and \textit{legal issue}, respectively. Based on these triplets, two text-attributed case graphs are constructed as the legal fact graph $G_{c,\text{fact}}=(V_{c,\text{fact}}, E_{c,\text{fact}})$ and the legal issue graph $G_{c,\text{issue}}=(V_{c,\text{issue}}, E_{c,\text{issue}})$. In $G_{c,\text{fact}}$, the node set $V_{c,\text{fact}}$ consists of all head and tail entities $h$ and $t$ in $R_{c,\text{fact}}$ represented as nodes $v_h$ and $v_t$. The edge set $E_{c,\text{fact}}$ includes directed edges $e_{v_hv_t}$ corresponding to relations $r$ from each head entity $h$ to its associated tail entity $t$. The construction of $G_{c,\text{issue}}$ follows the same procedure using $R_{c,\text{issue}}$. in $R_{c,\text{fact}}$ as $e_{v_hv_t}$. The same construction process is applied to the $G_{c,\text{issue}}$ for the legal issue. Additionally, each case’s fact graph and issue graph include a virtual global node $v_{\text{g, fact}}$ and $v_{\text{g, issue}}$ representing overall textual semantics, connected to all nodes to facilitate global information propagation.

\section{Method}
In this section, we present the LEXA framework in detail. The method begins by fine-tuning a retrieval oriented LLM embedding model for informative case representation generation, which is described in Section~\ref{sec:llm_fintuning}. With the proposed LEXA model, the node features are initialised through legal-domain prompts that guide LEXA to encode entities, relations, facts, and issues into text embeddings with rich semantic information, as described in Section~\ref{sec:llm}. A Legal Edge Updated Graph Neural Network (LEUGNN), proposed in Section~\ref{sec:LEUGNN}, refines these embeddings through message passing over legal relations within the case graph. To enhance generalization, graph augmentation is leveraged to construct diverse positive and negative samples for contrastive learning, as described in Section~\ref{sec:aug}. Finally, a contrastive learning objective, detailed in Section~\ref{sec:obj}, aligns embeddings of augmented views while preserving discriminative information, resulting in robust and semantically rich legal case representations.

\begin{figure}[!t]
\centering
\includegraphics[width=\textwidth]{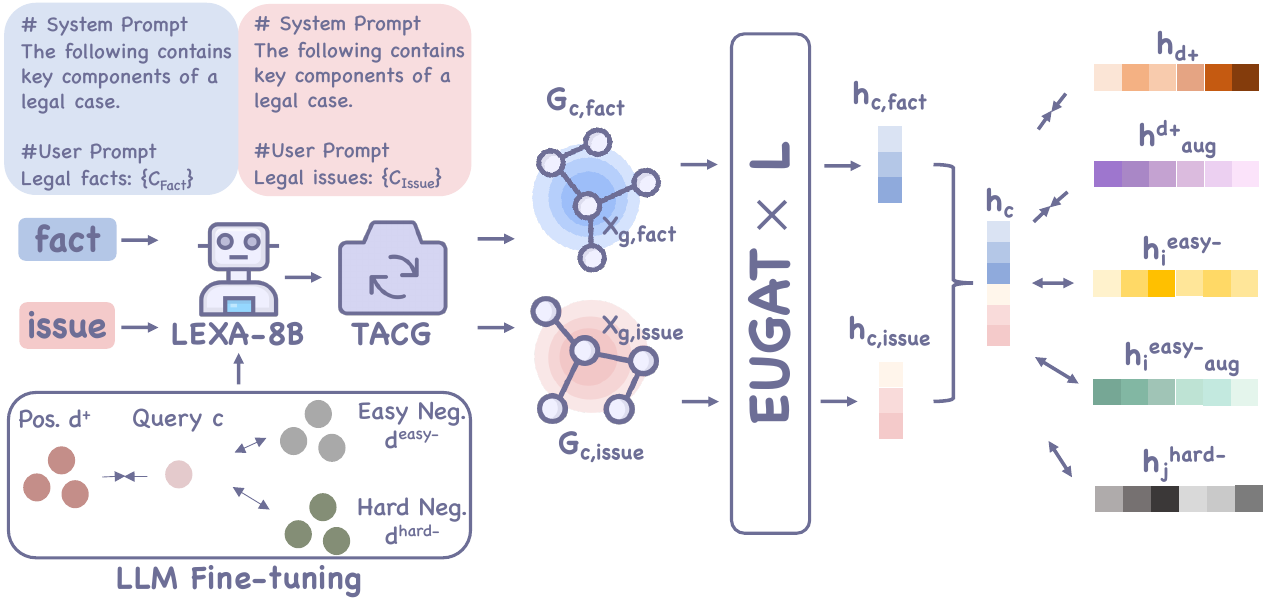}
\caption{Given a legal case $c$, the legal fact and legal issue sections are extracted and transformed into structured case graphs using LEXA-8B and case graph generation (TACG). Each case graph is encoded by L stacked Edge-Updated Graph Attention Network (EUGAT) layers to model semantic and relational dependencies. A readout function aggregates node representations into graph-level embeddings, which are combined to form the final case representation $h_{c}$. The model is trained end-to-end with a contrastive learning objective incorporating positive, easy and hard negative, and augmented samples to improve discriminative power and robustness.}
\label{fig:LEXA}
\end{figure}

\subsection{LEXA-8B}
\label{sec:llm_fintuning}
To develop a legal retrieval oriented embedding model capable of encoding legal cases into informative representations, the Qwen3-Embedding-8B~\cite{qwen3embedding} model was fine-tuned using supervised contrastive learning. The resulting model, \textsc{LEXA-8B}, addresses the base model’s limitation in capturing domain-specific semantics within the legal domain. Each case is first transformed into a structured text input integrating legal facts and legal issues. Through this fine-tuning process, \textsc{LEXA-8B} learns to align embeddings of relevant cases while distinguishing irrelevant ones, enabling it to capture fine-grained semantic relationships and generalise effectively across diverse legal scenarios, thus achieving precise and robust precedent retrieval.

\subsubsection{Contextualised Case Encoding}
During training, each case is reformulated as $c_\text{Cont}$ by concatenating its extracted legal facts and legal issues:

\begin{center}
\begin{encode}[title={Prompt Template for Contextualised Case Encoding}, label=prompt:recon]
\vspace{-1mm}
\hspace*{2em}\ \texttt{\# System Prompt}\\
\hspace*{2em}\ \texttt{The following contains key components of a legal case.} \\ \\
\hspace*{2em}\ \texttt{\# User Prompt}\\
\hspace*{2em}\ \texttt{Legal facts:} $\{c_{\text{Fact}}\}$. \\ 
\hspace*{2em}\ \texttt{Legal issues:} $\{c_{\text{Issue}}\}$. \\ 
\end{encode}
\end{center}

Feeding the reformulated case $c_\text{Cont}$ into \textsc{LEXA-8B}, the case is encoded into a dense vector representation as:
\begin{equation}
\label{eq:case encoding}
    \mathbf{x}_{c}=\text{\textsc{LEXA-8B}}(c_\text{Cont}),
\end{equation}
where $\mathbf{x}_{c}\in\mathbb{R}^d$ denotes the $d$-dimensional semantic embedding of the case, capturing its key legal element information for downstream retrieval task.

\subsubsection{LEXA-8B Fine-tuning Objective}
The supervised LLM Fine-tuning objective encourages embeddings of the query case and positive cases to be pulled closer together while pushing embeddings of negative cases apart, thereby enabling the model to learn discriminative and semantically enriched case representations. To further enhance training effectiveness, hard negative samples are introduced by retrieving cases with high BM25 relevance scores that are not labelled as ground-truth matches, ensuring the model is challenged with difficult but informative distinctions:
\begin{align}
\label{eq:cl_llm}
  \ell = -\text{log}\frac{e^{(s(\mathbf{x}_q,\mathbf{x}_{d^+}))/\tau}}{e^{(s(\mathbf{x}_q,\mathbf{x}_{d^+}))/\tau}+\sum\limits_{i=1}^{n} e^{(s(\mathbf{x}_q,\mathbf{x}_{d^{easy-}_i}))/\tau}+\sum\limits^m_{j=1}e^{(s(\mathbf{x}_q,\mathbf{x}_{d^{hard-}_j}))/\tau}}.
\end{align}
In Equation~\ref{eq:cl_llm}, $q$ denotes the query case. The positive sample $d^+$ is derived from the ground-truth labels. Negative samples are composed of two types: easy negatives $d^{easy-}$, which are randomly drawn from the candidate pool and include in-batch samples; and hard negatives $d^{hard-}$, which are retrieved using high BM25 similarity scores. The variables $n$ and $m$ denote the number of easy and hard negative samples, respectively. The similarity function $s$, implemented using either the dot product or cosine similarity, measures the similarities between case representations. Finally, the temperature coefficient $\tau$ scales the similarity scores, controlling the sharpness of the softmax distribution during contrastive learning.

\subsection{Prompt-based Graph Features Initialization with Large Language Models}
\label{sec:llm}
To generate graph features that capture richer contextual information from legal cases, node and edge features are encoded using large language models (LLMs) to enhance case representation and improving model training. Specifically, prompt is used in text encoding for guiding LLMs to focus on the legal task. 

\subsubsection{Prompt Templates.}
Recent research on prompt-based methods for legal case retrieval~\cite{promptcase} demonstrates that incorporating prompts can guide the model to capture richer semantic information, thereby improving performance on downstream tasks. Motivated by this, prompt-based encoding is adopted in feature extraction process. Given an example of relation triplet, ``The applicant'', ``is'', ``a Canadian'' and the legal facts and legal issues of the case, the prompts are shown in Table~\ref{tab:prompt_triplets}. A detail experiment of effectiveness of different prompt templates is described in Section~\ref{effet_llm}.
\begin{table*}[!t]\centering
\caption{Prompt templates for relation triplets features encoding.}\label{tab:prompt_triplets}
\resizebox{1\linewidth}{!}{
\begin{tabular}{lll}
\toprule
Type &Text &Prompt\\\midrule
\textbf{Relation Triplets} \\
\multirow{2}{*}{$\text{prompt}_{(\text{head})}$} &\multirow{2}{*}{``The applicant''} &``Given the following sentence, The applicant is a Canadian, \\
&&the head entity of legal relation triplet is: The applicant''\\\midrule
\multirow{2}{*}{$\text{prompt}_{(\text{tail})}$} &\multirow{2}{*}{``a Canadian''} &``Given the following sentence,The applicant is a Canadian, \\
&&the tail entity of legal relation triplet is: a Canadian''\\\midrule
\multirow{2}{*}{$\text{prompt}_{(\text{relation}})$} &\multirow{2}{*}{``is''} &``Given the following sentence, The applicant is a Canadian, \\
&&the relation between two entities of legal relation triplet is: is''\\\midrule
\textbf{Global Nodes} \\
$\text{prompt}_{(\text{fact})}$ &Legal facts text &``Legal facts: '' + Legal facts text\\\midrule
$\text{prompt}_{(\text{issue})}$ &Legal issues text &``Legal issues: '' + Legal issues text\\
\bottomrule
\end{tabular}}
\end{table*}

\subsubsection{Node and Edge Features of Relation Triplet.}
For a head entity node $v_{head}$, tail entity node $v_{tail}$, and relation edge between head and tail entity edge $e_{relation}$ in the case graph, the text encoding is conducted as follows:
\begin{align}
\label{eq:TAencoding}
    \mathbf{x}_{v_{head}} &= \text{LLM}(\text{prompt}_{(\text{head})});\\
    \mathbf{x}_{v_{tail}} &= \text{LLM}(\text{prompt}_{(\text{tail})});\\
    \mathbf{x}_{e_{relation}} &= \text{LLM}(\text{prompt}_{(\text{relation})}),
\end{align}
where $\text{prompt}_{(\text{head})}$, $\text{prompt}_{(\text{tail})}$, and $\text{prompt}_{(\text{relation})}$ are the encoding prompt of node $v_{head}$, node $v_{tail}$ and edge $e_{relation}$ respectively. LLM is a pre-trained text embedding large language model, such as text-embedding-3-large~\footnote{https://platform.openai.com/docs/guides/embeddings}, E5-Mistral-7B-Instruct~\cite{e5mistral} or Qwen3-Embedding-8B~\cite{qwen3embedding}. The vectors $\textbf{x}_{v_{head}} \in \mathbb{R}^d$, $\textbf{x}_{v_{tail}} \in \mathbb{R}^d$, and $\textbf{x}_{e_{relation}} \in \mathbb{R}^d$ represent the output of LLM encoding and serve as the feature representations in case graph for node $v_{head}$, $v_{tail}$ and edge $e_{relation}$. 

\subsubsection{Node and Edge Features of Virtual Global Node.}
For node features of the virtual global nodes $v_\text{g,fact}$ and $v_\text{g,issue}$, the whole texts of legal fact and legal issue are also encoded by the same LLM with prompt. To simplify the feature extraction process, the edge features connecting the global node to other nodes such as node $v$ are directly reused from the feature representation of node $v$:
\begin{align}
\label{eq:global_text}
    \mathbf{x}_{v_\text{g,fact}}&=\text{LLM}(\text{prompt}_{(\text{fact})});\\
    \mathbf{x}_{e_{\text{g,fact}}}&=\mathbf{x}_v;\\ 
    \mathbf{x}_{v_\text{g,issue}}&=\text{LLM}(\text{prompt}_{(\text{issue})});\\
    \mathbf{x}_{e_{\text{g,issue}}}&=\mathbf{x}_v.
\end{align}

\subsection{Legal Edge Updated Graph Neural Network}
\label{sec:LEUGNN}
Considering the importance of edges of case graph that represents the pivotal relations between legal entities, a novel Legal Edge Updated Graph Neural Network (LEUGNN) is proposed to incorporate the renewable nodes and edges features simultaneously. The LEUGNN includes two parts, the Edge-updated Graph Attention Layer (EUGAT) and readout function. 

\subsubsection{Edge-updated Graph Attention Layer.}
Edge-updated graph attention layer (EUGAT) is a fundamental component of LEUGNN for transferring the case graph into informative case representation, which includes different node feature update and edge feature update schemes respectively. 

$\bullet$\textbf{Edge Feature Update.} In EUGAT, the edge features are updated in every GNN layer instead of only updating the node features in CaseGNN. For the update of edge $e_{uv}$ that between node $u$ and node $v$ with $K$ attention heads is defined as:\begin{equation}
  \mathbf{h}^{'}_{e_{uv}} = \mathbf{h}_{e_{uv}}+\mathop{\text{Avg}}_{k=1:K}(\text{LeakyReLU}({\mathbf{w}_{\text{att,e}}^{k}}^{T} [\mathbf{W}^{k}_{n} \cdot \mathbf{h}_v \mathbin\Vert \mathbf{W}^{k}_{e} \cdot \mathbf{h}_{e_{uv}}\mathbin\Vert \mathbf{W}^{k}_{n} \cdot \mathbf{h}_u])),
\end{equation}
where $\mathbf{h}^{'}_{e_{uv}}\in\mathbb{R}^{d'}$ is the updated edge feature, and Avg means the average of the output vectors of K heads. All the weight matrices $\textbf{W}$ are in $\mathbb{R}^{d'\times d}$. Specifically, $\textbf{W}_{n}$ is the neighbour node update weight matrix and $\textbf{W}_{e}$ is the edge update weight matrix that are the same matrices used in Equation~\ref{eq:node_updat} and Equation~\ref{eq:node_attention}, $\mathbf{w}_{\text{att,e}}^{k} \in \mathbb{R}^{3d'}$ is the edge weight vector of attention layer. Moreover, to avoid over-smoothing problem, a residual connection is added. 

$\bullet$ \textbf{Node Feature Update.} For node feature update, a self-attention module will be used to aggregate the nodes and its neighbour nodes and edges information to an informative representation. Also, a residual connection is added for avoiding over-smoothing. As previous study shows, multi-head attention has better performance than original attention\cite{transformer}. According to multi-head attention mechanism, the update of node $v$ with $K$ attention heads is defined as:
\begin{equation}
\label{eq:node_updat}
  \mathbf{h}^{'}_v = \mathbf{h}_v + \mathop{\text{Avg}}_{k=1:K}(\sum_{u\in N(v)}\alpha^{k}(\mathbf{W}^{k}_n \cdot \mathbf{h}_u+\mathbf{W}^{k}_e \cdot \mathbf{h}_{e_{uv}})),
\end{equation}
where $\mathbf{h}^{'}_v\in\mathbb{R}^{d'}$ is the updated node feature, and Avg means the average of the output vectors of K heads. Specially, the input of the first EUGAT layer is initialised as $\textbf{h}_v=\textbf{x}_v$, $\textbf{h}_u=\textbf{x}_u$ and $\textbf{h}_{e_{uv}}=\textbf{x}_{e_{uv}}$. All the weight matrices $\textbf{W}$ are in $\mathbb{R}^{d'\times d}$. Specifically, $\textbf{W}_{n}$ is the neighbour node update weight matrix and $\textbf{W}_{e}$ is the edge update weight matrix respectively. $\alpha^{k}$ is the attention weight in the attention layer as:
\begin{equation}
\label{eq:node_attention}
    \alpha^{k}= \text{Softmax}(\text{LeakyReLU}({\mathbf{w}_{\text{att,n}}^{k}}^{T} [\mathbf{W}^{k}_{n} \cdot \mathbf{h}_v\mathbin\Vert \mathbf{W}^{k}_{n} \cdot \mathbf{h}_u \mathbin\Vert \mathbf{W}^{k}_{e} \cdot \mathbf{h}_{e_{uv}}])),
\end{equation}
where $\text{Softmax}$ is the softmax function, $\text{LeakyReLU}$ is the non-linear function, $\mathbf{w}_{\text{att,n}}^{k} \in \mathbb{R}^{3d'}$ is the node weight vector of attention layer and $\mathbin\Vert$ denotes concatenation of vectors. 

\subsubsection{Readout Function for Legal Case Representation}
\label{sec:Readout}
To generate the graph representation with the updated node and edge features, the Readout function is utilised in many GNN works~\cite{GIN,casegnn}, which is always defined as:
\begin{align}
\label{eq:readout}
  \mathbf{h}_{G} &= \text{Readout}(h_{v}^{\text(final)}|v \in G),
\end{align}
where $h_{v}^{\text(final)} \in \mathbb{R}^{d'}$ denotes the representation of node $v$ in graph $G$ and Readout is an aggregation function that produces a graph-level representation through operations such as max pooling, sum pooling, or attention pooling. 

Specifically, in this paper, since the virtual global is connected to all other nodes in the graph, the final iteration feature of the virtual global node $\mathbf{h}_{v_g} \in \mathbb{R}^{d'}$ can be regarded as the final representation of the case graph due to its full graph connectivity. Accordingly, in the design of LEUGNN, the Readout function in Equation~\ref{eq:readout} is implemented by directly using the representation of the global nodes $\mathbf{h}_{v_{g,\text{fact}}} \in \mathbb{R}^{d'}$ and $\mathbf{h}_{v_{g,\text{issue}}} \in \mathbb{R}^{d'}$ respectively:
\begin{align}
  \mathbf{h}_{c,\text{fact}} &= \mathbf{h}_{v_{g,\text{fact}}},\\
  \mathbf{h}_{c,\text{issue}} &= \mathbf{h}_{v_{g,\text{issue}}}.
\end{align}

The final representation of case $c$, $\textbf{h}_c \in \mathbb{R}^{2d'}$ is the concatenation of legal fact graph representation $\textbf{h}_{c,\text{fact}}$ and legal issue graph representation $\textbf{h}_{c,\text{issue}}$, which is defined as:
\begin{equation}
  \textbf{h}_c = \textbf{h}_{c,\text{fact}} \mathbin\Vert \textbf{h}_{c,\text{issue}}.
\end{equation}

\subsection{Graph Augmentation}
\label{sec:aug}
Graph augmentation is utilised in this paper to generate more positive and negative samples for graph contrastive learning. To effectively obtain the augmented graph while reserving the informative legal structure information, the graph augmentation methods of edge dropping and feature masking are leveraged for graph contrastive learning during the training process.

\subsubsection{Edge Dropping} Edge dropping is a popular graph augmentation method that randomly drop the edges of the original graph as a new graph. Given a case graph $G_c$, the edge dropping augmentation graph is denoted as $G_{c,\text{drop}}$, where the adjacency matrices of $G_c$ and $G_{c,\text{drop}}$ is defined as:
\begin{equation}
  \mathbf{A}_{c,\text{drop}} = \mathbf{M} \odot \mathbf{A}_{c},
\end{equation}
where $\textbf{A}_{c} \in \mathbb{R}^{n\times n}$ is the adjacency matrix of the graph $G_c$. $\textbf{M} \in {\{0,1\}}^{n\times n}$ is a binary mask matrix that acts on the adjacency matrix for edge dropping, which generates the adjacency matrix $\mathbf{A}_{c,\text{drop}}\in \mathbb{R}^{n\times n}$ for augmented graph. Specifically, $\textbf{M}_{i,j}=\text{Bernoulli}(\epsilon)$, where $\epsilon \in (0,1)$ refers to the drop rate hyperparameter and $\odot$ indicates the Hadamard product (also known as element-wise product).

\subsubsection{Feature Masking}
Another popular graph augmentation method called feature masking is also leveraged for training in this paper. Specifically, to mask the node and edge features in a graph, the column of node and edge feature tensors will be masked in a probability $p_{\text{node}}\in (0,1)$ and $p_{\text{edge}}\in (0,1)$ respectively, which are two hyperparameters for training.

\subsection{Graph Augmentation Contrastive Learning Objective Function}
\label{sec:obj}
Contrastive learning is a tool that aims at pulling the positive samples closer while pushing the negative samples far away to provide the training signal for training~\cite{SAILER,MVCL,CL4LJP}. The goal of training the LEXA model is to distinguish the relevant cases from irrelevant cases given a large case pool, which corresponds to the goal of contrastive learning. In contrast to CaseGNN, LEXA incorporates augmented graphs into the contrastive learning objective to enrich the training signals and enhance representation learning. Accordingly, the objective function for model training is defined as follows: 
\begin{small}
\begin{align}
\label{eq:cl}
  &\ell = -\text{log}\\\notag
  &\left(\frac{e^{\frac{s(\mathbf{h}_q,\mathbf{h}_{d^+})}{\tau}}+e^{\frac{s(\mathbf{h}_q,\mathbf{h}_{d_{aug}^+})}{\tau}}}{e^{\frac{s(\mathbf{h}_q,\mathbf{h}_{d^+})}{\tau}}+e^{\frac{s(\mathbf{h}_q,\mathbf{h}_{d_{aug}^+})}{\tau}}+\sum\limits_{i=1}^{n} e^{\frac{s(\mathbf{h}_q,\mathbf{h}_{d^{easy-}_i})}{\tau}}+\sum\limits^n_{i=1}e^{\frac{s(\mathbf{h}_q,\mathbf{h}_{d^{easy-}_{i,aug}})}{\tau}}+\sum\limits^m_{j=1}e^{\frac{s(\mathbf{h}_q,\mathbf{h}_{d^{hard-}_j})}{\tau}}}\right).
\end{align}
\end{small}
In equation~\ref{eq:cl}, $q$ denotes the query case. The positive sample $d^+$ is directly derived from the ground truth labels in the dataset. Easy negative samples $d^{easy-}$ are drawn randomly from the entire candidate pool and also include in-batch samples from other queries. To further enhance training effectiveness, hard negative samples $d^{hard-}$ are incorporated by identifying challenging cases that are difficult to distinguish from relevant ones. A candidate case is considered a hard negative if it receives a high BM25 relevance score but is not labelled as a ground truth match. Such cases exhibit high textual similarity to the query while remaining irrelevant, thereby offering more informative contrastive signals. The variables $n$ and $m$ represent the number of easy and hard negative samples, respectively.  The similarity function $s$, which measures the closeness between representations, can be implemented using either the dot product or cosine similarity. The temperature coefficient $\tau$ is used to scale the similarity scores, controlling the sharpness of the softmax distribution during contrastive learning.

Specifically, in the LEXA model, graph augmentation is applied to both positive and negative cases to provide additional training signals, addressing the challenges posed by limited and costly legal data annotation. In Equation~\ref{eq:cl}, the augmented graph of relevant cases $d^+$ is denoted as $d_{aug}^+$ while the augmented graph of easy negative sample $d^{easy-}$ is denoted as $d_{aug}^{easy-}$. For each case graph, only a single augmentation strategy is employed to generate an augmented graph. This decision is motivated by the empirical finding that using multiple augmentation methods simultaneously results in reduced performance. Additionally, during training, either augmented positive samples or augmented easy negative samples are used individually, as combining both was found to yield suboptimal results. 

\section{Experiments}
In this section, the following research questions (RQs) will be studied extensively to evaluate the effectiveness of the proposed LEXA model:
\begin{itemize}
    \item RQ1: How does LEXA perform compared with other legal case retrieval models?
    \item RQ2: How effective are edge feature update and graph augmentation in LEXA?
    \item RQ3: How effective is EUGAT compared with other GNNs?
    \item RQ4: How different LLMs affect the performance
    \item RQ5: How effective is prompt-based LLM features encoding?
    \item RQ6: How does graph contrastive learning help with LCR in LEXA?
    \item RQ7: How does graph sparsity affect retrieval performance in LEXA?
    \item RQ8: How do hyper-parameter settings affect LEXA?
\end{itemize}

\subsection{Setup}
\label{sec:setup}
In setup, the detailed descriptions of two benchmark datasets, seven evaluated metrics, seven LCR baselines and the experiment implementations are provided. 

\subsubsection{Datasets.}
To assess the effectiveness of the proposed LEXA model, experiments are conducted on two benchmark legal case retrieval datasets, COLIEE2022~\cite{COLIEE2022} and COLIEE2023~\cite{COLIEE2023}, both released as part of the Competition on Legal Information Extraction and Entailment (COLIEE). These datasets consist of legal cases sourced from the Federal Court of Canada. The task of legal case retrieval is retrieving relevant cases from the full candidate pool given a query case. Although the training sets of COLIEE2022 and COLIEE2023 partially overlap, their test sets are entirely distinct. Additionally, as shown in Table~\ref{tab:dataset}, the average number of relevant cases per query differs between the two datasets, resulting in varying levels of retrieval difficulty. These datasets represent the most widely used English benchmarks for legal case retrieval. Notably, LEXA can be readily adapted to multilingual settings through appropriate legal information extraction tools and language models.

\begin{table*}[!t]
\centering
\caption{Statistics of datasets.}
\label{tab:dataset}
\resizebox{0.8\linewidth}{!}{
\begin{tabular}{c|cc|cc}
    \toprule
    \multirow{2}{*}{Datasets} &\multicolumn{2}{c|}{COLIEE2022} &\multicolumn{2}{c}{COLIEE2023} \\
    \cmidrule{2-5}
    &train &test &train &test \\\midrule
    \# Query &898 &300 &959 &319 \\
    \# Candidates &4415 &1563 &4400 &1335 \\
    \# Avg. relevant cases &4.68 &4.21 &4.68 &2.69 \\
    Avg. length (\# token) &6724 &6785 &6532 &5566 \\
    Largest length (\# token) &127934 &85136 &127934 &61965 \\
    \bottomrule
\end{tabular}}
\end{table*}

\subsubsection{Baselines.}
\label{baselines}
According to the recent research~\cite{promptcase,SAILER}, six popular and state-of-the-art methods are compared as well as the competition winners:
\begin{itemize}
    \item \textbf{BM25}~\cite{BM25} is a classic term-matching model that remains a strong competitor in many retrieval tasks. Building on the TF–IDF framework~\cite{TF-IDF}, BM25 incorporates document length normalization and a corpus-level length prior to mitigate biases arising from varying document sizes.
    \item \textbf{LEGAL-BERT}~\cite{LEGAL-BERT} is a bert-based model pre-trained on a large and diverse corpus of English legal text spanning legislation and judicial opinions from multiple jurisdictions. Its domain-specific pre-training makes it a strong benchmark for legal NLP tasks.
    \item \textbf{MonoT5}~\cite{monot5} adapts the T5 text-to-text architecture~\cite{T5} to ranking scenarios. By framing relevance estimation as a generation problem, MonoT5 demonstrates powerful language modelling capabilities in retrieval settings.
    \item \textbf{SAILER}~\cite{SAILER} is a structure-aware legal language model designed specifically for legal case retrieval. Its encoder–decoder architecture is trained to capture the internal structure of cases, enabling it to achieve competitive retrieval performance. 
    \item \textbf{PromptCase}~\cite{promptcase} is a prompt-based reformulation method that guides language models to focus on key legal aspects. By explicitly encoding legal facts and issues, it enhances the model’s ability to retrieve cases that match the key legal elements of a query.
    \item \textbf{E5-Mistral-7B-Instruct}~\cite{e5mistral}is a general purpose embedding model built through instruction tuning on a broad set of corpora. It produces strong semantic embeddings and is widely used for retrieval and semantic search tasks. 
    \item \textbf{Qwen3-Embedding-8B}~\cite{qwen3embedding} generates high-quality representations suitable for similarity and retrieval applications. Its extensive training and robust generalization make it a reliable baseline even in specialized domains such as law.
    \item \textbf{Inf-Retriever-V1}~\cite{inf-retriever-v1} is optimized for large-scale retrieval workloads, focusing on creating highly discriminative embeddings that balance retrieval accuracy with computational efficiency.
    \item \textbf{CaseGNN}~\cite{casegnn} models each case as a graph enriched with textual attributes, using a graph neural network and contrastive learning to derive case representations. It serves as an important reference point for graph-based methods for legal case retrieval task.
    \item \textbf{CaseLink}~\cite{caselink} advances legal case retrieval by incorporating connections among cases. By leveraging inter-case links, it captures latent legal dependencies that further enhance retrieval quality.
\end{itemize}

\subsubsection{Metrics.}
\label{metrics}
In this experiment, seven metrics that often utilised for evaluating information retrieval models and LCR models are leveraged to comprehensively evaluate the LEXA model. According to the previous LCR works~\cite{promptcase,LeCaRD,SAILER}, top 5 ranking results are evaluated and all metrics are the higher the better. The details of the seven metrics are as follows:
\begin{itemize}
    \item \textbf{Precision (P)} captures how reliably a model identifies relevant items. It is defined as: $\text{Precision} ={\text{TP}}/({\text{TP}+\text{FP}})$, where TP represents correctly retrieved relevant cases, and FP counts the cases incorrectly retrieved as relevant.
    \item \textbf{Recall (R)} evaluates the extent to which a model successfully retrieves all relevant items. It is calculated by $\text{Recall}=\text{TP}/(\text{TP}+\text{FN})$, with FN denoting relevant cases the model failed to retrieve.
    \item \textbf{Micro F1 (Mi-F1)} offers a unified measure that balances precision and recall by computing their harmonic mean across all instances: $\text{Mi-F1}=2\times(\text{Precision}\times \text{Recall})/(\text{Precision} + \text{Recall}))$. 
    \item \textbf{Macro F1 (Ma-F1)} assesses performance in a more class-agnostic manner by averaging the F1 score of each class, ensuring equal contribution regardless of class size: $\text{Ma-F1}=\frac{1}{N}\sum_{i=1}^N\text{F1}_i$.
    \item \textbf{Mean Reciprocal Rank (MRR) @K} measures how quickly a system retrieves the first relevant item. It is defined as $\text{MRR} = \frac{1}{N}\sum_{i=1}^N\frac{1}{\text{rank}_i}$, where $N$ is the number of queries, $\text{rank}_i$ is the rank position of the first correct result for the $i$-th query.  
    \item \textbf{Mean Average Precision (MAP) @K} captures the average retrieval precision across all queries. For each query, average precision is computed separately, and then aggregated as $\text{MAP} = \frac{1}{N}\sum_{i=1}^N\text{AP}_i$.
    \item \textbf{Normalized Discounted Cumulative Gain (NDCG) @K} examines ranking quality by rewarding relevant items appearing near the top while penalizing those placed lower. This measure is computed as $\text{NDCG@K} = \text{DCG@K}/\text{IDCG@K}$, where DCG captures the discounted gain from the actual ranked list, and IDCG represents the best possible ordering within the top K positions.
\end{itemize} 

\subsubsection{Implementation.}
\label{implementation}
The implementation includes data preprocessing and hyperparameters.  

$\bullet$\textbf{Data Processing.} All French text is removed from both datasets. The relation and entity extraction required for case graph construction is performed using the spaCy\footnote{\url{https://spacy.io/}}, Stanford OpenIE~\cite{OpenIE} and LexNLP\footnote{\url{https://github.com/LexPredict/lexpredict-lexnlp}} toolkits. 

$\bullet$\textbf{Hyperparameters.} The number of EUGAT layers are set to 2 and the number of EUGAT heads are chosen from \{1, 2, 4\}. The training batch sizes are chosen from \{16, 32, 64, 128\}. The Dropout~\cite{dropout} rate of every layer's representation is chosen from \{0.1, 0.2, 0.3, 0.4, 0.5\}. Adam~\cite{Adam} is applied as optimiser with the learning rate chosen from \{0.00001, 0.00005, 0.0001, 0.0005, 0.000005\} and weight decay from \{0.00001, 0.0001, 0.001\}. During training, for every query, the number of positive sample is set to 1; the number of randomly chosen easy negative sample is set to 1; the number of hard negative samples is chosen from \{0, 1, 5\}. The in-batch samples from other queries are also employed as easy negative samples. Qwen3-Embedding-8B\cite{qwen3embedding} is chosen as the LLM model for features generation. The embedding size of the features are set to 4096. The augmentation probabilities of $\epsilon$, $p_{\text{node}}$ and $p_{\text{edge}}$ in Section~\ref{sec:aug} are chosen from \{0, 0.1, 0.3, 0.5\}. Only augmented positive samples or augmented easy negative samples will be utilised, as using both augmented positive and random negative samples together results in poor performance. Two-stage experiment uses the top 10 retrieved cases by BM25 as the first stage result. 

\begin{table*}[!t]\centering
\caption{Overall performance on COLIEE2022 (\%). Underlined numbers indicate the best baselines. Bold numbers indicate the best performance of all methods. Both one-stage and two-stage Top-5 results are reported.}\label{tab:overall_2022}
\resizebox{1\linewidth}{!}{
\begin{tabular}{c|l|ccccccc}
\toprule
\midrule
&\multirow{2}{*}{Methods} &\multicolumn{7}{c}{COLIEE2022} \\
\cmidrule{3-9}
&&P@5 &R@5 &Mi-F1 &Ma-F1 &MRR@5 &MAP &NDCG@5 \\\midrule
\midrule
&\textbf{One-stage}\\
&BM25 &17.9 &21.2 &19.4 &21.4 &23.6 &25.4 &33.6\\
\midrule
\multirow{4}{*}{\rotatebox{90}{LM}}&LEGAL-BERT &4.47 &5.30 &4.85 &5.38 &7.42 &7.47 &10.9\\
&MonoT5 &0.71 &0.65 &0.60 &0.79 &1.39 &1.41 &1.73\\
&SAILER &16.6 &15.2 &14.0 &16.8 &17.2 &18.5 &25.1\\
&PromptCase &17.1 &20.3 &18.5 &20.5 &35.1 &33.9 &38.7\\
\midrule
\multirow{3}{*}{\rotatebox{90}{LLM}}&E5-Mistral-7B-Instruct &21.4 &25.4 &23.2 &25.7 &26.8 &28.5 &38.0\\
&Qwen3-Embedding-8B &21.6 &25.7 &23.5 &26.0 &26.7 &29.0 &37.8\\
&Inf-Retriever-V1 &21.1 &25.1 &22.9 &25.5 &26.6 &28.7 &37.9 \\
\midrule
\multirow{3}{*}{\rotatebox{90}{Graph}}&CaseGNN&35.5 $\pm$0.2 &42.1 $\pm$0.2 &38.4 $\pm$0.3 &42.4 $\pm$0.1 &66.8 $\pm$0.8 &64.4 $\pm$0.9 &69.3 $\pm$0.8\\
&CaseLink &\underline{37.0} $\pm$0.1 &\underline{43.9} $\pm$0.1 &\underline{40.1} $\pm$0.1 &\underline{44.2} $\pm$0.1 &\underline{67.3} $\pm$0.5 &\underline{65.0} $\pm$0.2 &\underline{70.3} $\pm$0.1\\
&\cellcolor{lightgray}LEXA (Ours) &\cellcolor{lightgray}43.7 $\pm$0.2 &\cellcolor{lightgray}51.9 $\pm$0.2 &\cellcolor{lightgray}47.5 $\pm$0.2 &\cellcolor{lightgray}52.4 $\pm$0.3 &\cellcolor{lightgray}77.0 $\pm$0.5 &\cellcolor{lightgray}74.7 $\pm$0.9 &\cellcolor{lightgray}79.3 $\pm$0.8 \\
\midrule
&\textbf{Two-stage} \\
\multirow{2}{*}{\rotatebox{90}{LM}}&SAILER &23.8 &25.7 &24.7 &25.2 &43.9 &42.7 &48.4 \\
&PromptCase &23.5 &25.3 &24.4 &\textbf{\underline{30.3}} &41.2 &39.6 &45.1\\
\midrule
\multirow{3}{*}{\rotatebox{90}{LLM}}&E5-Mistral-7B-Instruct &22.0 &26.2 &24.0 &25.8 &48.4 &46.8 &51.5 \\
&Qwen3-Embedding-8B &21.9 &26.0 &23.8 &25.8 &48.8 &48.0 &52.1\\
&Inf-Retriever-V1 &22.2 &26.4 &24.1 &26.1 &45.5 &44.5 &49.6\\
\midrule
\multirow{3}{*}{\rotatebox{90}{Graph}}&CaseGNN &22.9 $\pm$0.1 &27.2 $\pm$0.1 &24.9 $\pm$0.1 &27.0 $\pm$0.1 &54.9 $\pm$0.4 &54.0 $\pm$0.5 &57.3 $\pm$0.6\\
&CaseLink &\underline{24.7} $\pm$0.1 &\underline{29.1} $\pm$0.1 &\underline{26.8} $\pm$0.1 &29.2 $\pm$0.1 &\underline{56.0} $\pm$0.2 &\underline{55.0} $\pm$0.2 &\underline{58.6} $\pm$0.1\\
&\cellcolor{lightgray}LEXA (Ours) &\cellcolor{lightgray}25.4 $\pm$0.2 &\cellcolor{lightgray}30.2 $\pm$0.2 &\cellcolor{lightgray}27.6 $\pm$0.2 &\cellcolor{lightgray}30.0 $\pm$0.1 &\cellcolor{lightgray}59.0 $\pm$0.6 &\cellcolor{lightgray}58.0 $\pm$0.8 &\cellcolor{lightgray}61.2 $\pm$0.6\\
\bottomrule
\end{tabular}}
\end{table*}

\begin{table*}[!t]\centering
\caption{Overall performance on COLIEE2023 (\%). Underlined numbers indicate the best baselines. Bold numbers indicate the best performance of all methods. Both one-stage and two-stage Top-5 results are reported.}\label{tab:overall_2023}
\resizebox{1\linewidth}{!}{
\begin{tabular}{c|l|cccccccc}
\toprule
\midrule
&\multirow{2}{*}{Methods} &\multicolumn{7}{c}{COLIEE2023} \\
\cmidrule{3-9}
&&P@5 &R@5 &Mi-F1 &Ma-F1 &MRR@5 &MAP &NDCG@5 \\\midrule
\midrule
\multirow{2}{*}{\rotatebox{90}{ }}&\textbf{One-stage}\\
&BM25 &16.5 &30.6 &21.4 &22.2 &23.1 &20.4 &23.7\\
\midrule
\multirow{4}{*}{\rotatebox{90}{LM}} &LEGAL-BERT &4.64 &8.61 &6.03 &6.03 &11.4 &11.3 &13.6\\
&MonoT5 &0.38 &0.70 &0.49 &0.47 &1.17 &1.33 &0.61 \\
&SAILER &12.8 &23.7 &16.6 &17.0 &25.9 &25.3 &29.3\\
&PromptCase &16.0 &29.7 &20.8 &21.5 &32.7 &32.0 &36.2 \\
\midrule
\multirow{3}{*}{\rotatebox{90}{LLM}}&E5-Mistral-7b-Instruct &16.0 &29.7 &20.8 &21.5 &21.9 &22.9 &30.8 \\
&Qwen3-Embedding-8B &18.4 &34.1 &23.9 &25.2 &25.8 &27.5 &36.4\\
&Inf-Retriever-V1 &19.0 &35.3 &24.7 &26.0 &26.5 &28.0 &37.6 \\
\midrule
\multirow{3}{*}{\rotatebox{90}{Graph}}&CaseGNN &17.7 $\pm$0.7 &32.8 $\pm$0.7&23.0 $\pm$0.5 &23.6 $\pm$0.5 &38.9 $\pm$1.1 &37.7 $\pm$0.8 &42.8 $\pm$0.7\\
&CaseLink &\underline{20.9} $\pm$0.3&\underline{38.4} $\pm$0.6&\underline{27.1} $\pm$0.3&\underline{28.2} $\pm$0.3&\underline{45.8} $\pm$0.5&\underline{44.3} $\pm$0.7&\underline{49.8} $\pm$0.4\\
&\cellcolor{lightgray}LEXA &\cellcolor{lightgray}22.6 $\pm$0.4 &\cellcolor{lightgray}41.9 $\pm$0.8 &\cellcolor{lightgray}29.3 $\pm$0.6 &\cellcolor{lightgray}30.5 $\pm$0.7 &\cellcolor{lightgray}48.7 $\pm$0.7 &\cellcolor{lightgray}47.0 $\pm$0.7 &\cellcolor{lightgray}52.6 $\pm$0.8\\
\midrule
&\textbf{Two-stage} \\
\multirow{2}{*}{\rotatebox{90}{LM}} &SAILER &19.6 &32.6 &24.5 &23.5 &37.3 &36.1 &40.8\\
&PromptCase &\textbf{\underline{21.8}} &36.3 &\underline{27.2} &26.5 &39.9 &38.7 &44.0\\
\midrule
\multirow{3}{*}{\rotatebox{90}{LLM}} &E5-Mistral-7b-Instruct &19.7 &36.6 &25.6 &26.5 &46.9 &44.4 &49.9\\
&Qwen3-Embedding-8B &20.1 &37.4 &26.2 &27.0 &47.7 &46.3 &51.1\\
&Inf-Retriever-V1 &19.7 &36.6 &25.6 &26.6 &48.2 &46.9 &51.6\\
\midrule
\multirow{3}{*}{\rotatebox{90}{Graph}} &CaseGNN &20.2 $\pm$0.2 &37.6 $\pm$0.5 &26.3 $\pm$0.3 &27.3 $\pm$0.2 &45.8 $\pm$0.9 &44.4 $\pm$0.8 &49.6 $\pm$0.8\\
&CaseLink &21.0 $\pm$0.3 &\underline{38.9} $\pm$0.5 &27.1 $\pm$0.3 &\underline{28.2} $\pm$0.3 &\underline{48.8} $\pm$0.2 &\underline{47.2} $\pm$0.1 &\underline{52.6} $\pm$0.1\\
&\cellcolor{lightgray}LEXA &\cellcolor{lightgray}21.3 $\pm$0.2 &\cellcolor{lightgray}39.5 $\pm$0.3 &\cellcolor{lightgray}27.7 $\pm$0.2 &\cellcolor{lightgray}28.6 $\pm$0.2 &\cellcolor{lightgray}50.6 $\pm$0.2 &\cellcolor{lightgray}49.3 $\pm$0.2 &\cellcolor{lightgray}54.1 $\pm$0.2\\
\bottomrule
\end{tabular}}
\end{table*}

\subsection{Overall Performance (RQ1)}
\label{overall}
The results in Table~\ref{tab:overall_2022} and Table~\ref{tab:overall_2023} reveal clear performance patterns across both datasets. Traditional retrieval methods and language model baselines occupy the lower end of the accuracy performance, while methods that incorporate structural information especially graph-based models achieve substantially better retrieval quality. Across every evaluation setting, LEXA delivers the strongest results and consistently establishes new state of the art performance.

On COLIEE2022, BM25 and language model based systems such as LEGAL-BERT, MonoT5, SAILER, and PromptCase achieve modest accuracy and weak ranking performance. The gap between these models and large language model embeddings widens considerably. E5-Mistral-7B-Instruct, Qwen3-Embedding-8B, and Inf-Retriever-V1 provide a noticeable boost, surpassing all language model baselines by large margins. However, none of these embedding based methods come close to matching the performance of the graph-based models. CaseGNN and CaseLink demonstrate substantial improvements over language model and large language model methods, underscoring the advantage of explicitly modelling legal entities and relations. LEXA further extends this advantage, outperforming CaseLink by large absolute margins. These gains indicate a significant advancement in retrieval capability.

A similar trend appears on COLIEE2023. The language model and large language model baselines again show limited ability to capture the structural reasoning required for legal case retrieval. In contrast, graph-based models dominate the upper end of the performance distribution. CaseLink continues to be the strongest baseline on this dataset across nearly all metrics. Yet LEXA surpasses CaseLink on every measure by a consistent and meaningful amount, setting new records in both precision oriented and ranking oriented metrics.

The two stage retrieval pipeline further highlights the relative strengths of the different approaches. While language model and large language model methods benefit moderately from using BM25 as a first stage filter, their improvements remain limited. Graph-based approaches, however, show consistent gains especially on COLIEE2023 where the candidate set provided by BM25 is more informative. Even in scenarios where CaseLink narrows the gap, LEXA remains the strongest method across all metrics and both datasets. The improvements observed in the two-stage setting confirm that LEXA is effective not only in an end-to-end retrieval scenario but also when functioning as a re-ranking model.

Overall, the experimental results demonstrate that methods relying solely on lexical similarity or sentence level semantics are insufficient for capturing the complex relational structure of legal cases. Graph-based models provide a clear advantage, and LEXA by combining prompt-guided initialization of node representations with graph message passing and contrastive graph augmentation achieves the highest accuracy, recall, and ranking performance among all evaluated methods.

\begin{table*}[!t]\centering
\caption{Ablation study. (\%)}\label{tab:ablation}
\resizebox{1\linewidth}{!}{
\begin{tabular}{l|ccccccc}
\toprule
\multirow{2}{*}{Methods} &\multicolumn{7}{c}{COLIEE2022}\\
\cmidrule{2-8}
&P@5 &R@5 &Mi-F1 &Ma-F1 &MRR@5 &MAP &NDCG@5\\
\midrule
PromptCase &17.1 &20.3 &18.5 &20.5 &35.1 &33.9 &38.7\\
w/o $v_g$ &1.6 $\pm$0.1 &2.9 $\pm$0.1 &2.1 $\pm$0.1 &2.2 $\pm$0.1 &4 $\pm$0.1 &4.0 $\pm$0.1 &4.8 $\pm$0.2\\
Avg Readout &30.5 $\pm$0.5 &36.2 $\pm$0.6 &33.1 $\pm$0.5 &36.8 $\pm$0.5 &61.3 $\pm$0.4 &59.0 $\pm$0.1 &64.6 $\pm$0.5\\
CaseGNN &35.5 $\pm$0.2 &42.1 $\pm$0.2 &38.4 $\pm$0.3 &42.4 $\pm$0.1 &66.8 $\pm$0.8 &64.4 $\pm$0.9 &69.3 $\pm$0.8 \\
+GCL &35.5 $\pm$0.3 &42.2 $\pm$0.4 &38.6 $\pm$0.3 &42.5 $\pm$0.6 &67.1 $\pm$0.7 &64.6 $\pm$0.1 &70.2 $\pm$0.5\\
+EUGAT &35.9 $\pm$0.5 &42.7 $\pm$0.6 &39.2 $\pm$0.7 &43.3 $\pm$0.5 &67.2 $\pm$0.9 &64.0 $\pm$1.2 &69.4 $\pm$1.2\\
w/o Tuned LLM &36.5 $\pm$0.6 &43.3 $\pm$0.7 &39.6 $\pm$0.6 &43.8 $\pm$0.7 &68.1 $\pm$1.1 &65.3 $\pm$1.1 &70.8 $\pm$1.1\\
LEXA &\textbf{43.7} $\pm$0.2 &\textbf{51.9} $\pm$0.2 &\textbf{47.5} $\pm$0.2 &\textbf{52.4} $\pm$0.3 &\textbf{77.0} $\pm$0.5 &\textbf{74.7} $\pm$0.9 &\textbf{79.3} $\pm$0.8 \\
\midrule
\midrule
\multirow{2}{*}{Methods}&\multicolumn{7}{c}{COLIEE2023}\\
\cmidrule{2-8}
&P@5 &R@5 &Mi-F1 &Ma-F1 &MRR@5 &MAP &NDCG@5\\
\midrule
PromptCase &16.0 &29.7 &20.8 &21.5 &32.7 &32.0 &36.2\\
w/o $v_g$ &1.6 $\pm$0.1 &2.9 $\pm$0.1 &2.1 $\pm$0.1 &2.2 $\pm$0.1 &4.0 $\pm$0.1 &2.9 $\pm$0.1 &4.8 $\pm$0.1\\
Avg Readout &17.6 $\pm$0.4 &32.6 $\pm$0.7 &22.8 $\pm$0.5 &23.6 $\pm$0.4 &37.7 $\pm$1.0 &36.6 $\pm$0.7 &41.7 $\pm$0.5 \\
CaseGNN &17.7 $\pm$0.7 &32.8 $\pm$0.7&23.0 $\pm$0.5 &23.6 $\pm$0.5 &38.9 $\pm$1.1 &37.7 $\pm$0.8 &42.8 $\pm$0.7\\
+GCL &17.7 $\pm$0.1 &32.9 $\pm$0.2 &23.0 $\pm$0.1 &23.7 $\pm$0.2 &39.1 $\pm$0.3 &37.8 $\pm$0.2 &42.9 $\pm$0.2\\
+EUGAT &18.0 $\pm$0.3 &33.4 $\pm$0.5 &23.4 $\pm$0.3 &24.0 $\pm$0.2 &39.3 $\pm$0.7 &38.1 $\pm$0.7 &43.2 $\pm$0.5\\
w/o Tuned LLM &18.2 $\pm$0.3 &33.8 $\pm$0.4 &23.7 $\pm$0.4 &24.3 $\pm$0.3 &40.0 $\pm$0.2 &38.9 $\pm$0.3 &43.8 $\pm$0.3\\
LEXA &\textbf{22.6} $\pm$0.4 &\textbf{41.9} $\pm$0.8 &\textbf{29.3} $\pm$0.6 &\textbf{30.5} $\pm$0.7 &\textbf{48.7} $\pm$0.7 &\textbf{47.0} $\pm$0.7 &\textbf{52.6} $\pm$0.8\\
\bottomrule
\end{tabular}}
\end{table*}

\subsection{Ablation Study (RQ2)}
The ablation analysis in Table~\ref{tab:ablation} evaluates a series of progressively enhanced model variants in order to understand how each design choice contributes to LEXA.  (1) PromptCase, a purely text based setting without graph; (2) using the case graph that excludes the virtual global node (w/o $v_g$); (3) using the average pooling in CaseGNN (Avg Readout); (4) CaseGNN; (5) extending CaseGNN by incorporating graph contrastive learning (+GCL); (6) changing CaseGNN GNN layer from EdgeGAT to EUGAT (+EUGAT); (7) The full LEXA system combines both enhancements in a unified framework. All variants are tested on the two datasets using every evaluation measure in a single stage retrieval setting.

The ablation results in Table~\ref{tab:ablation} provide a clear picture of how each component contributes to the performance of LEXA. Across both COLIEE2022 and COLIEE2023, the progression from simple text based representations to fully structured graph learning highlights a consistent trend: the richer the modelling of case structure and node interactions, the stronger the retrieval ability. PromptCase offers a text only baseline that relies solely on the encoded fact and issue statements. While reasonably strong for a non-graph method, its performance is far below any configuration that incorporates graph structure. Removing the global node leads to a dramatic collapse in effectiveness on both datasets. The near zero values across all metrics indicate that graph message passing without a meaningful semantic anchor prevents the model from forming any coherent case representation. Introducing even a simple aggregation mechanism substantially shifts the landscape. The Avg Readout variant, which pools node embeddings by averaging, provides a sizeable jump compared to PromptCase on both datasets. This confirms that the structural cues present in the case graphs carry essential information, even when used in a minimal way. Enhancements to CaseGNN further highlight the importance of more expressive graph reasoning. Adding contrastive learning (+GCL) consistently improves ranking metrics such as MRR@5, MAP, and NDCG@5, showing that the additional training objective provides stronger supervision for distinguishing subtle differences between similar cases. Replacing EdgeGAT with EUGAT (+EUGAT) yields another set of gains, reflecting the benefits of more flexible and informative edge updates. The improvements are visible across nearly all metrics in both dataset. The variant that removes the tuned large language model (w/o Tuned LLM) sits between the enhanced CaseGNN variants and the full LEXA model. Although it performs competitively, its results show that high quality node initialization remains important, especially for capturing fine-grained semantic distinctions that downstream graph layers build upon.

At the top of the comparison, LEXA achieves the strongest performance by combining all proposed components, including EUGAT and graph contrastive learning. LEXA outperforms every ablated configuration by large margins. For instance, on COLIEE2022 it surpasses the best non-LEXA variant by more than 9\% in NDCG@5, and the improvements on COLIEE2023 are similarly substantial. These consistent gains demonstrate that the combination of expressive graph modelling, semantically grounded global representations, and contrastive training produces a significantly more informative case representation.

\begin{table*}[!t]\centering
\caption{Effectiveness of GNNs. (\%)}\label{tab:effect_gnn}
\resizebox{1\linewidth}{!}{
\begin{tabular}{l|ccccccc}
\toprule
\multirow{2}{*}{Methods} &\multicolumn{7}{c}{COLIEE2022}\\
\cmidrule{2-8}
&P@5 &R@5 &Mi-F1 &Ma-F1 &MRR@5 &MAP &NDCG@5 \\
\midrule
GCN &21.3 $\pm$0.3 &25.3 $\pm$0.4 &23.2 $\pm$0.2 &26.0 $\pm$0.4 &46.0 $\pm$0.2 &44.4 $\pm$0.1 &49.7 $\pm$0.3\\
GAT &29.3 $\pm$0.1 &34.8 $\pm$0.3 &31.8 $\pm$0.1 &35.3 $\pm$0.2 &59.2 $\pm$0.5 &56.9 $\pm$0.3 &62.3 $\pm$0.7\\
EdgeGAT &35.5 $\pm$0.2 &42.1 $\pm$0.2 &38.4 $\pm$0.3 &42.4 $\pm$0.1 &66.8 $\pm$0.8 &64.4 $\pm$0.9 &69.3 $\pm$0.8\\
EUGAT &\textbf{43.7} $\pm$0.2 &\textbf{51.9} $\pm$0.2 &\textbf{47.5} $\pm$0.2 &\textbf{52.4} $\pm$0.3 &\textbf{77.0} $\pm$0.5 &\textbf{74.7} $\pm$0.9 &\textbf{79.3} $\pm$0.8\\
\midrule
\midrule
\multirow{2}{*}{Methods}&\multicolumn{7}{c}{COLIEE2023}\\
\cmidrule{2-8}
&P@5 &R@5 &Mi-F1 &Ma-F1 &MRR@5 &MAP &NDCG@5\\\midrule
GCN &12.8 $\pm$0.2 &23.7 $\pm$0.3 &16.5 $\pm$0.1 &16.9 $\pm$0.2 &27.8 $\pm$0.8 &26.8 $\pm$0.6 &31.6 $\pm$0.7\\
GAT &17.4 $\pm$0.3 &32.2 $\pm$0.5 &22.5 $\pm$0.3 &23.1 $\pm$0.5 &37.5 $\pm$0.4 &36.4 $\pm$0.4 &41.4 $\pm$0.4\\
EdgeGAT &17.7 $\pm$0.7 &32.8 $\pm$0.7 &23.0 $\pm$0.5 &23.6 $\pm$0.5 &38.9 $\pm$1.1 &37.7 $\pm$0.8 &42.8 $\pm$0.7\\
EUGAT &\textbf{22.6} $\pm$0.4 &\textbf{41.9} $\pm$0.8 &\textbf{29.3} $\pm$0.6 &\textbf{30.5} $\pm$0.7 &\textbf{48.7} $\pm$0.7 &\textbf{47.0} $\pm$0.7 &\textbf{52.6} $\pm$0.8\\
\bottomrule
\end{tabular}}
\end{table*}

\subsection{Effectiveness of GNNs (RQ3)}
To investigate the contribution of the EUGAT layer, EUGAT is replaced with several commonly used GNN layers of GCN~\cite{GCN}, GAT~\cite{GAT}, and EdgeGAT. Results on both datasets across all evaluation metrics are summarized in Table~\ref{tab:effect_gnn}.

EUGAT distinguishes itself from previous graph attention mechanisms through its layer-wise edge update strategy. While EdgeGAT incorporates edge features when updating node embeddings, it does not revise those edge features across layers. In contrast, EUGAT continuously updates edge representations, enabling the model to iteratively refine the semantic information carried by each relation. This capability is crucial for legal case graphs, where edges encode the nuanced legal relationships between entities. By progressively enhancing these relation embeddings, EUGAT equips LEXA with more expressive and discriminative edge representations. As a result, the model can better capture the relational patterns that determine case categories, ultimately improving its ability to retrieve the most relevant cases for a given query.

\subsection{Effectiveness of Different LLM for Feature Encoding (RQ4)}
\begin{table*}[!t]\centering
\caption{Performance of different LLMs used for feature encoding on COLIEE2022 and COLIEE2023 (\%).}\label{tab:llm}
\resizebox{1\linewidth}{!}{
\begin{tabular}{l|ccccccc}
\toprule
\multirow{2}{*}{Methods} &\multicolumn{7}{c}{COLIEE2022}\\
\cmidrule{2-8}
&P@5 &R@5 &Mi-F1 &Ma-F1 &MRR@5 &MAP &NDCG@5\\\midrule
E5-Mistral-7B-Instruct &41.8 $\pm$0.7 &50.1 $\pm$0.4 &45.8 $\pm$0.3 &50.6 $\pm$0.5 &73.7 $\pm$0.2 &71.1 $\pm$0.1 &76.5 $\pm$0.2 \\
Qwen3-Embedding-8B &43.2 $\pm$0.3 &51.3 $\pm$0.4 &46.9 $\pm$0.4 &51.6 $\pm$0.3 &75.8 $\pm$0.2 &73.5 $\pm$0.2 &78.0 $\pm$0.2 \\
LEXA &\textbf{43.7} $\pm$0.2 &\textbf{51.9} $\pm$0.2 &\textbf{47.5} $\pm$0.2 &\textbf{52.4} $\pm$0.3 &\textbf{77.0} $\pm$0.5 &\textbf{74.7} $\pm$0.9 &\textbf{79.3} $\pm$0.8 \\
\midrule
\toprule
\multirow{2}{*}{Methods} &\multicolumn{7}{c}{COLIEE2023}\\
\cmidrule{2-8}
&P@5 &R@5 &Mi-F1 &Ma-F1 &MRR@5 &MAP &NDCG@5\\\midrule
E5-Mistral-7B-Instruct &20.9 $\pm$0.1 &38.7 $\pm$0.1 &27.1 $\pm$0.1 &28.1 $\pm$0.1 &45.1 $\pm$0.2 &43.6 $\pm$0.1 &49.4 $\pm$0.1\\
Qwen3-Embedding-8B &22.2 $\pm$0.1 &41.2 $\pm$0.1 &28.8 $\pm$0.1 &29.9 $\pm$0.1 &46.9 $\pm$0.1 &45.6 $\pm$0.3 &51.2 $\pm$0.1\\
LEXA &\textbf{22.6} $\pm$0.4 &\textbf{41.9} $\pm$0.8 &\textbf{29.3} $\pm$0.6 &\textbf{30.5} $\pm$0.7 &\textbf{48.7} $\pm$0.7 &\textbf{47.0} $\pm$0.7 &\textbf{52.6} $\pm$0.8\\
\bottomrule
\end{tabular}}
\end{table*}
In this section, the effectiveness of various LLMs for feature encoding is evaluated on both
datasets across all evaluation metrics, with results reported in Table~\ref{tab:llm}. Both E5-Mistral-7B-Instruct and Qwen3-Embedding-8B are employed to encode relation triplets into node and edge feature embeddings. Specifically, Qwen3-Embedding-8B is adopted as the feature encoder in LEXA.

On COLIEE2022, the improvements are relatively modest but consistent. For example, Qwen3-Embedding-8B achieves a higher MAP and NDCG@5, reflecting stronger ranking quality. Similar marginal gains are observed in precision, recall, and F1 scores, indicating that Qwen3-Embedding-8B provides more effective embeddings for both node and edge features. On COLIEE2023, the advantage of Qwen3-Embedding-8B becomes more pronounced. It achieves 2.5\% improvement in R@5, leading to higher micro-F1 and macro-F1. The improvements in MAP and NDCG@5 also suggest that Qwen3-Embedding-8B yields more reliable retrieval rankings. Given that COLIEE2023 is a more challenging dataset, these consistent gains highlight the robustness and generalization capability of Qwen3-Embedding-8B.

\subsection{Effectiveness of Prompt-based LLM Feature Encoding (RQ5)}
\label{effet_llm}
\begin{table*}[!t]\centering
\caption{Prompt templates for relation triplets features encoding.}\label{tab:effect_prompt}
\resizebox{1\linewidth}{!}{
\begin{tabular}{l|ll}
\toprule
 &Type &Prompt Templates\\\midrule
\multirow{9}{*}{prompt 0} &$\text{prompt}_{(\text{head})}$ &Given the following sentence, The applicant is a Canadian,\\
&&the head entity of legal relation triplet is: The applicant\\
\cmidrule{2-3}
&$\text{prompt}_{(\text{tail})}$ &Given the following sentence, The applicant is a Canadian,\\
&&the tail entity of legal relation triplet is: a Canadian\\
\cmidrule{2-3}
&$\text{prompt}_{(\text{relation})}$  &Given the following sentence, The applicant is a Canadian,\\
&&the relation between two entities of legal relation triplet is: is\\
\cmidrule{2-3}
&$\text{prompt}_{(\text{fact})}$  &Legal facts:  \\
\cmidrule{2-3}
&$\text{prompt}_{(\text{issue})}$  &Legal issues:  \\\midrule
\multirow{15}{*}{prompt 1} &$\text{prompt}_{(\text{head})}$ &You are a legal knowledge extraction system tasked with constructing\\
&&knowledge graphs from legal texts. Please identify and encode the following \\
&&head entity that captures the meaningful legal relationships in the case:\\
\cmidrule{2-3}
&$\text{prompt}_{(\text{tail})}$ &You are a legal knowledge extraction system tasked with constructing\\
&&knowledge graphs from legal texts. Please identify and encode the following\\
&&tail entity that captures the meaningful legal relationships in the case:\\
\cmidrule{2-3}
&$\text{prompt}_{(\text{relation})}$  &You are a legal knowledge extraction system tasked with constructing\\
&&knowledge graphs from legal texts. Please identify and encode the following \\
&&relations that captures the meaningful legal relationships in the case:\\
\cmidrule{2-3}
&$\text{prompt}_{(\text{fact})}$  &You are a legal assistant trained to extract and encode structured legal\\
&&information from case texts, identify and encode the following legal fact: \\
\cmidrule{2-3}
&$\text{prompt}_{(\text{issue})}$  &You are a legal assistant trained to extract and encode structured legal\\
&&information from case texts, identify and encode the following legal issue:  \\\midrule
\multirow{10}{*}{prompt 2} &$\text{prompt}_{(\text{head})}$ &Given the following sentence, The applicant is a Canadian,\\
&&the head entity of legal relation triplet is:\\
\cmidrule{2-3}
&$\text{prompt}_{(\text{tail})}$ &Given the following sentence, The applicant is a Canadian,\\
&&the tail entity of legal relation triplet is:\\
\cmidrule{2-3}
&$\text{prompt}_{(\text{relation})}$  &Given the following sentence, The applicant is a Canadian,\\
&&the relation between two entities of legal relation triplet is\\
\cmidrule{2-3}
&$\text{prompt}_{(\text{fact})}$  &Legal facts: \\
\cmidrule{2-3}
&$\text{prompt}_{(\text{issue})}$  &Legal issues:  \\\midrule
\multirow{10}{*}{prompt 3} &$\text{prompt}_{(\text{head})}$ &Instruction: Encode the legal head entity, The applicant,\\
&&from the following legal sentence: The applicant is a Canadian.\\
\cmidrule{2-3}
&$\text{prompt}_{(\text{tail})}$ &Instruction: Encode the legal tail entity, a Canadian,\\
&&from the following legal sentence: The applicant is a Canadian.\\
\cmidrule{2-3}
&$\text{prompt}_{(\text{relation})}$  &Instruction: Encode the legal relation, is,\\
&&from the following legal sentence: The applicant is a Canadian.\\
\cmidrule{2-3}
&$\text{prompt}_{(\text{fact})}$  &Legal facts: \\
\cmidrule{2-3}
&$\text{prompt}_{(\text{issue})}$  &Legal issues:  \\
\bottomrule
\end{tabular}}
\end{table*}

\begin{table*}[!t]\centering
\caption{Effectiveness of different prompt templates. Details of prompt templates are shown in Table~\ref{tab:effect_prompt}
}\label{tab:prompt}
\resizebox{1\linewidth}{!}{
\begin{tabular}{l|ccccccc}
\toprule
\multirow{2}{*}{Methods} &\multicolumn{7}{c}{COLIEE2022}\\
\cmidrule{2-8}
&P@5 &R@5 &Mi-F1 &Ma-F1 &MRR@5 &MAP &NDCG@5\\\midrule
no prompt &39.8 $\pm$1.1 &47.3 $\pm$1.3 &43.3 $\pm$1.2 &47.9 $\pm$1.4 &71.6 $\pm$2.2 &69.1 $\pm$2.1 &74.7 $\pm$2.0\\
prompt 0 &43.7 $\pm$0.2 &51.9 $\pm$0.2 &47.5 $\pm$0.2 &52.4 $\pm$0.3 &77.0 $\pm$0.5 &74.7 $\pm$0.9 &79.3 $\pm$0.8 \\
prompt 1 &41.3 $\pm$0.9 &49.0 $\pm$1.0  &44.8 $\pm$1.0 &49.4 $\pm$1.1 &74.0 $\pm$0.6 &71.6 $\pm$0.2 &76.6 $\pm$0.3\\
prompt 2 &41.8 $\pm$0.6 &49.6 $\pm$0.7 &45.4 $\pm$0.6 &50.1 $\pm$0.5 &74.7 $\pm$1.5 &71.9 $\pm$1.1 &77.2 $\pm$0.8 \\
prompt 3 &40.5 $\pm$1.7 &48.1 $\pm$2.0 &43.9 $\pm$1.8 &48.3 $\pm$2.0 &73.0 $\pm$2.6 &70.2 $\pm$2.3 &75.7 $\pm$1.9 \\\midrule
\toprule
\multirow{2}{*}{Methods} &\multicolumn{7}{c}{COLIEE2023}\\
\cmidrule{2-8}
&P@5 &R@5 &Mi-F1 &Ma-F1 &MRR@5 &MAP &NDCG@5\\\midrule
no prompt &22.0 $\pm$0.5 &40.8 $\pm$0.8 &28.6 $\pm$0.6 &29.6 $\pm$0.8 &47.1 $\pm$1.1 &45.5 $\pm$1.2 &51.0 $\pm$1.3\\
prompt 0 &22.6 $\pm$0.4 &41.9 $\pm$0.8 &29.3 $\pm$0.6 &30.5 $\pm$0.7 &48.7 $\pm$0.7 &47.0 $\pm$0.7 &52.6 $\pm$0.8 \\
prompt 1 &21.7 $\pm$0.6 &40.3  $\pm$1.1 &28.2 $\pm$0.8 &29.3  $\pm$0.9 &48.9 $\pm$0.5 &47.3  $\pm$0.5 &52.5 $\pm$0.9\\
prompt 2 &21.7 $\pm$0.2 &40.3 $\pm$0.4 &28.2 $\pm$0.3 &29.2 $\pm$0.4 &47.4 $\pm$1.5 &45.9 $\pm$1.4 &51.2 $\pm$1.1\\
prompt 3 &21.8 $\pm$0.5 &40.5 $\pm$0.9 &28.3 $\pm$0.6 &29.4 $\pm$0.7 &46.1 $\pm$0.2 &47.7 $\pm$0.3 &51.6 $\pm$0.3 \\
\bottomrule
\end{tabular}}
\end{table*}
This section examines the effectiveness of prompt-based LLM feature encoding in legal case retrieval. Firstly, the impact of alternative prompt templates is assessing on the COLIEE2022 and COLIEE2023 datasets. Given an example of relation triplet, `The applicant', `is', `a Canadian' and the legal facts and legal issues of the case, four prompt templates are shown in the Table~\ref{tab:effect_prompt}. And the performance of different prompt templates is shown in Table~\ref{tab:prompt}. 

On COLIEE2022, applying prompts leads to clear and consistent improvements over the no-prompt baseline across all evaluation metrics. Among the four templates, prompt 0 achieves the strongest performance overall. This suggests that a concise, direct, and minimally intrusive prompt is most effective for guiding LLMs to extract case relevant semantic features. Prompts 1 and 2 also improve over the no-prompt setting but fall slightly behind prompt 0, indicating that additional verbosity or narrative structure does not necessarily yield better retrieval representations. Prompt 3 underperforms relative to other templates, showing that not all forms of instructive prompting are equally beneficial for this task.

While on COLIEE2023, the impact of prompting is notably smaller. All prompt templates perform within a narrow margin. This suggests that, for this dataset, the LLM is less sensitive to prompt wording, likely due to dataset characteristics such as shorter cases or more straightforward relation structures that reduce reliance on prompt-induced guidance. Even so, prompt-based variants remain comparable to or slightly better than the baseline, demonstrating that prompts do not harm performance even when their benefits are limited. These results show the sensitivity of prompt templates and the importance of selecting concise and well structured templates.

\subsection{Effectiveness of Graph Augmentation for Graph Contrastive Learning (RQ6)}
This experiment evaluates the effectiveness of graph contrastive learning (GCL) by comparing different graph augmentation strategies including node feature masking, edge feature masking, and edge dropping on both COLIEE2022 and COLIEE2023. All methods are assessed under a one stage setting using seven retrieval metrics.
\begin{table*}[!t]\centering
\caption{Effectiveness of graph contrastive learning. (\%)}\label{tab:effect_gcl}
\resizebox{1\linewidth}{!}{
\begin{tabular}{l|ccccccc}
\toprule
\multirow{2}{*}{Methods} &\multicolumn{7}{c}{COLIEE2022}\\
\cmidrule{2-8}
&P@5 &R@5 &Mi-F1 &Ma-F1 &MRR@5 &MAP &NDCG@5 \\
\midrule
w/o GCL &35.9 $\pm$0.5 &42.7 $\pm$0.6 &39.2 $\pm$0.7 &43.3 $\pm$0.5 &67.2 $\pm$0.9 &64.0 $\pm$1.2 &69.4 $\pm$1.2\\
w/ FeatMask-Node &41.0 $\pm$0.6 &48.7 $\pm$0.8 &44.6 $\pm$0.7 &48.9 $\pm$0.9 &74.2 $\pm$1.4 &71.4 $\pm$1.2 &76.5 $\pm$1.4\\
w/ FeatMask-Edge &41.1 $\pm$1.1 &48.8 $\pm$1.3 &44.6 $\pm$1.1 &49.2 $\pm$1.3 &73.9 $\pm$1.4 &71.0 $\pm$1.3 &76.6 $\pm$1.1\\
w/ EdgeDrop &\textbf{43.7} $\pm$0.2 &\textbf{51.9} $\pm$0.2 &\textbf{47.5} $\pm$0.2 &\textbf{52.4} $\pm$0.3 &\textbf{77.0} $\pm$0.5 &\textbf{74.7} $\pm$0.9 &\textbf{79.3} $\pm$0.8\\
\midrule
\midrule
\multirow{2}{*}{Methods}&\multicolumn{7}{c}{COLIEE2023}\\
\cmidrule{2-8}
&P@5 &R@5 &Mi-F1 &Ma-F1 &MRR@5 &MAP &NDCG@5\\\midrule
w/o GCL &18.0 $\pm$0.3 &33.4 $\pm$0.5 &23.4 $\pm$0.3 &24.0 $\pm$0.2 &39.3 $\pm$0.7 &38.1 $\pm$0.7 &43.2 $\pm$0.5\\
w/ FeatMask-Node &22.2 $\pm$0.3 &41.3 $\pm$0.5 &28.9 $\pm$0.4 &30.0 $\pm$0.5 &48.0 $\pm$0.5 &46.2 $\pm$1.0 &52.0 $\pm$0.9 \\
w/ FeatMask-Edge &22.6 $\pm$0.3 &41.9 $\pm$0.6 &29.3 $\pm$0.4 &30.4 $\pm$0.5 &48.2 $\pm$1.1 &46.5 $\pm$1.0 &52.3 $\pm$1.0\\
w/ EdgeDrop &\textbf{22.6} $\pm$0.4 &\textbf{41.9} $\pm$0.8 &\textbf{29.3} $\pm$0.6 &\textbf{30.5} $\pm$0.7 &\textbf{48.7} $\pm$0.7 &\textbf{47.0} $\pm$0.7 &\textbf{52.6} $\pm$0.8\\
\bottomrule
\end{tabular}}
\end{table*}

As shown in Table~\ref{tab:effect_gcl}, applying GCL with any augmentation strategy yields substantial improvements over the model trained without GCL on both datasets, confirming that contrastive learning enhances the model’s ability to capture structural and semantic patterns in legal case graphs. Among the three augmentation methods, edge dropping consistently achieves the best performance, outperforming both feature masking strategies across all metrics and datasets. Unlike node or edge feature masking, which alters only the attribute space, edge dropping modifies the structural topology of the graph. This leads to larger perturbations between augmented views, enabling the contrastive objective to learn more discriminative and robust representations. Between the two feature masking augmentations, edge feature masking performs better than node feature masking on both datasets. Since edges in the case graph encode relations between legal entities which are central to legal reasoning and retrieval, masking edge attributes encourages the model to better capture and reconstruct relation semantics. This also aligns with the design of LEXA, which places emphasis on updating and utilizing edge features. The relative improvement of edge masking therefore highlights the importance of rich and well modelled edge representations for achieving higher retrieval accuracy.

\subsection{Impact of Graph Sparsification via Edge Pruning (RQ7)}

\begin{table*}[!t]\centering
\caption{Impact of graph sparsification via edge pruning.}\label{tab:graph_sparcity}
\resizebox{1\linewidth}{!}{
\begin{tabular}{l|ccccccc}
\toprule
\multirow{2}{*}{Methods} &\multicolumn{7}{c}{COLIEE2022}\\
\cmidrule{2-8}
&P@5 &R@5 &Mi-F1 &Ma-F1 &MRR@5 &MAP &NDCG@5\\\midrule
80\% Edge &39.6 $\pm$0.4 &47.1 $\pm$0.5 &43.0 $\pm$0.5 &47.7 $\pm$0.5 &71.0 $\pm$0.3 &68.4 $\pm$0.3 &74.5 $\pm$0.1\\
90\% Edge &40.5 $\pm$0.5 &48.1 $\pm$0.6 &43.9 $\pm$0.6 &48.6 $\pm$0.7 &72.6 $\pm$2.2 &70.0 $\pm$1.6 &75.7 $\pm$1.5 \\
100\% Edge &\textbf{43.7} $\pm$0.2 &\textbf{51.9} $\pm$0.2 &\textbf{47.5} $\pm$0.2 &\textbf{52.4} $\pm$0.3 &\textbf{77.0} $\pm$0.5 &\textbf{74.7} $\pm$0.9 &\textbf{79.3} $\pm$0.8\\\midrule
\toprule
\multirow{2}{*}{Methods} &\multicolumn{7}{c}{COLIEE2023}\\
\cmidrule{2-8}
&P@5 &R@5 &Mi-F1 &Ma-F1 &MRR@5 &MAP &NDCG@5\\\midrule
80\% Edge  &22.2 $\pm$0.1 &41.3  $\pm$0.1 &28.9 $\pm$0.1 &29.9 $\pm$0.1 &47.9 $\pm$0.3 &46.1 $\pm$0.3 &51.7 $\pm$0.3\\
90\% Edge &22.4 $\pm$0.4 &41.7 $\pm$0.2 &29.2 $\pm$0.2 &30.3 $\pm$0.2 &48.6 $\pm$0.4 &46.8 $\pm$0.6 &51.4 $\pm$0.5 \\
100\% Edge &\textbf{22.6} $\pm$0.4 &\textbf{41.9} $\pm$0.8 &\textbf{29.3} $\pm$0.6 &\textbf{30.5} $\pm$0.7 &\textbf{48.7} $\pm$0.7 &\textbf{47.0} $\pm$0.7 &\textbf{52.6} $\pm$0.8\\
\bottomrule
\end{tabular}}
\end{table*}

To evaluate how reducing graph density affects retrieval effectiveness, three experiments with different edge pruning ratios are conducted on both datasets, as shown in Table~\ref{tab:graph_sparcity}. Specifically, the 80\% edge setting retains 80\% of the edges from the original TACG graph, the 90\% edge setting retains 90\% of the edges, and the 100\% edge setting corresponds to the original, unpruned graph.
Overall, the results indicate that reducing graph density through edge pruning leads to a steady degradation in retrieval effectiveness across all evaluation metrics. The unpruned graph consistently achieves the best performance on both datasets, outperforming pruned variants in terms of P@5, F1, MAP, MRR@5, and NDCG@5. Moderate edge pruning (90\%) results in noticeable but limited performance drops, while more aggressive pruning (80\%) causes further degradation. This trend suggests that excessive edge removal disrupts important structural and semantic relationships that are essential for accurate legal case retrieval.
Although the performance gap between pruned and unpruned graphs is smaller on COLIEE2023 than on COLIEE2022, the fully connected graph remains consistently superior. These findings demonstrate that preserving rich graph connectivity is crucial for stable ranking quality across different dataset scales, and naive edge pruning introduces a clear accuracy–efficiency trade-off.

\subsection{Parameter Sensitivity (RQ8)}
\begin{figure}[ht]
\centering
\subfigure[]{
    \includegraphics[width=0.45\linewidth]{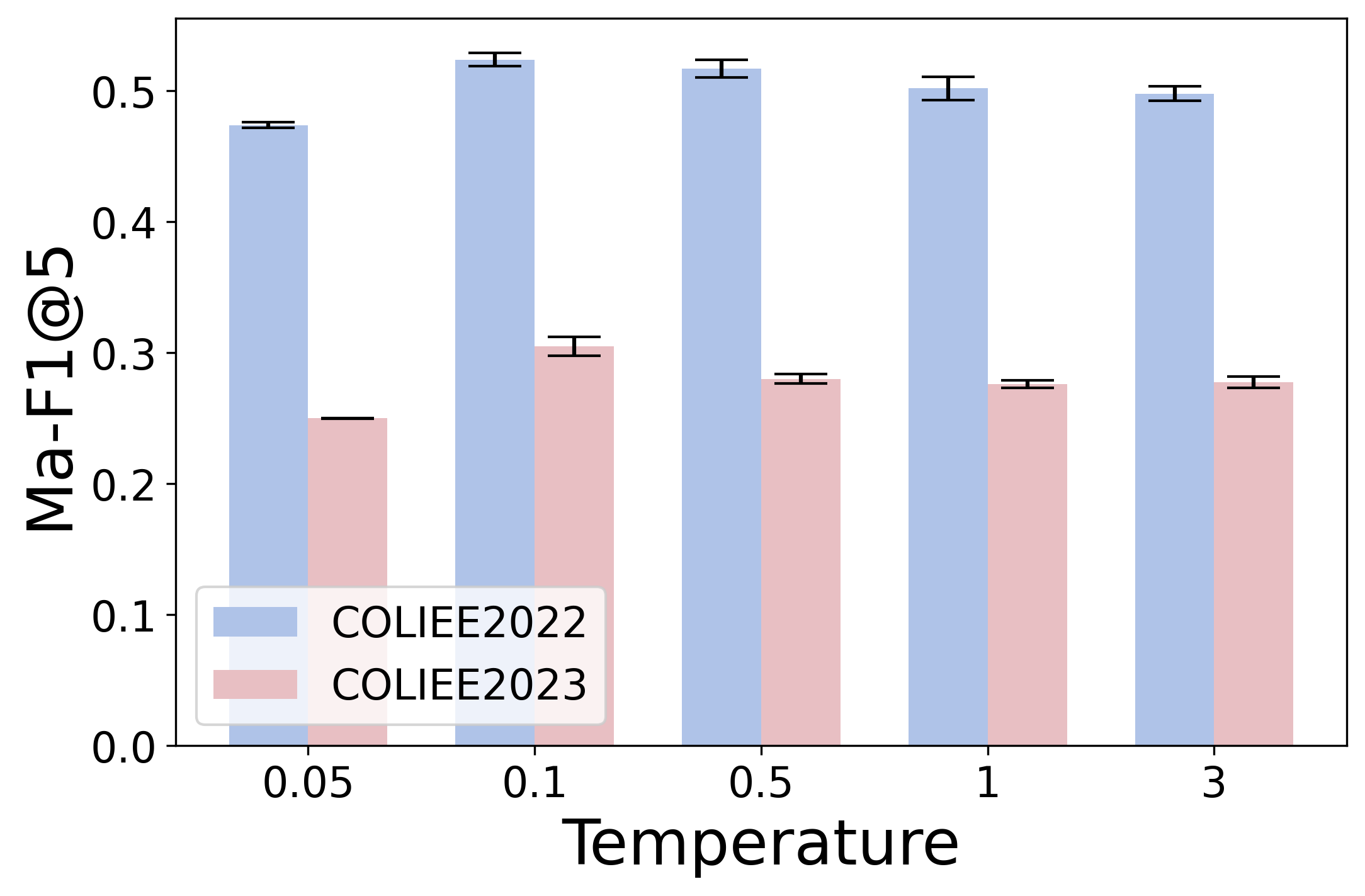}
    \label{fig:temp-maf1}
}
\hfill
\subfigure[]{
    \includegraphics[width=0.45\linewidth]{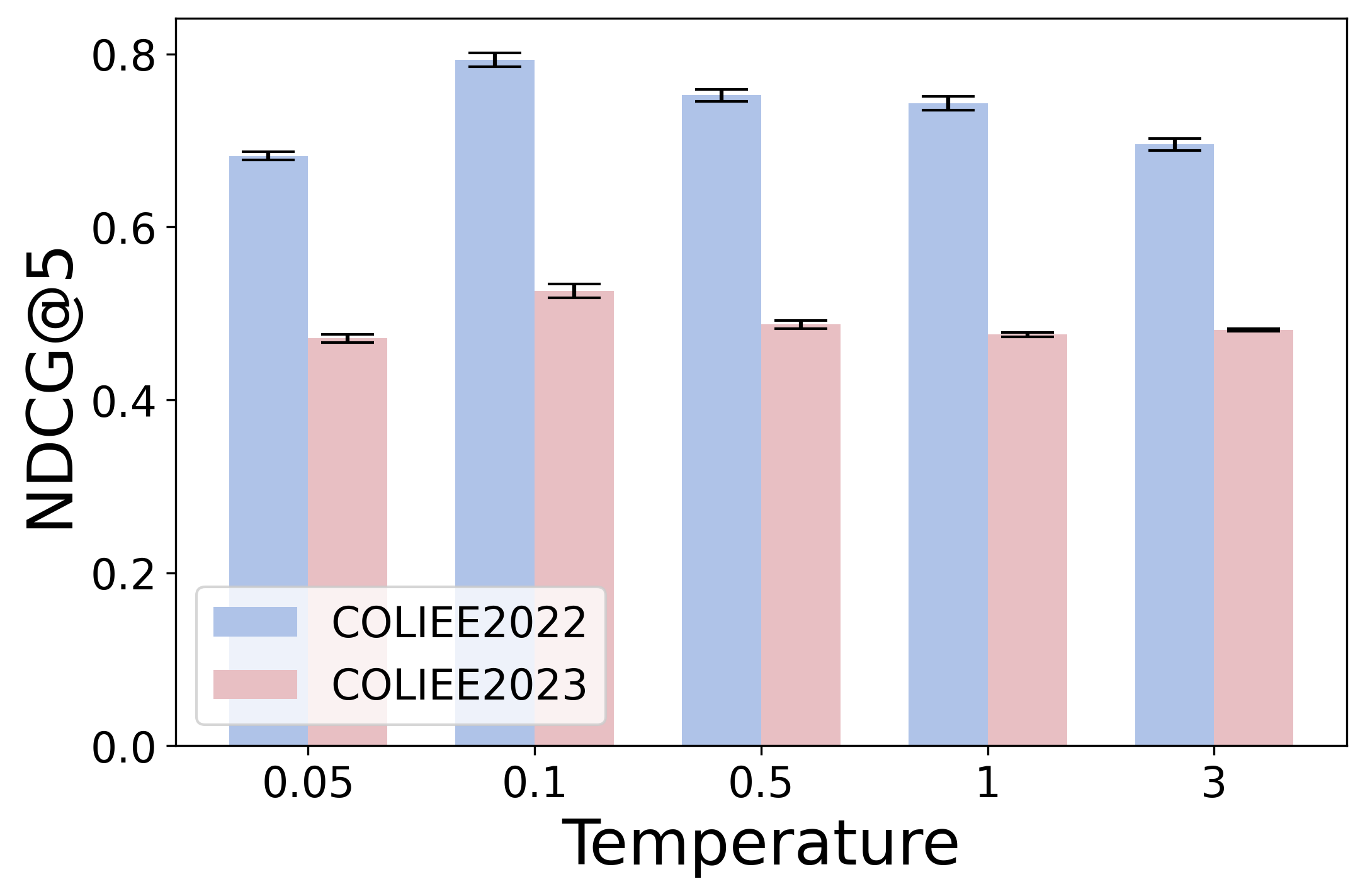}
    \label{fig:temp-ndcg}
}
\caption{Parameter sensitivity for the temperature $\tau$ in the contrastive loss of LEXA.}
\label{fig:temp}
\end{figure}

\begin{figure}[ht]
\centering
\subfigure[]{
    \includegraphics[width=0.45\linewidth]{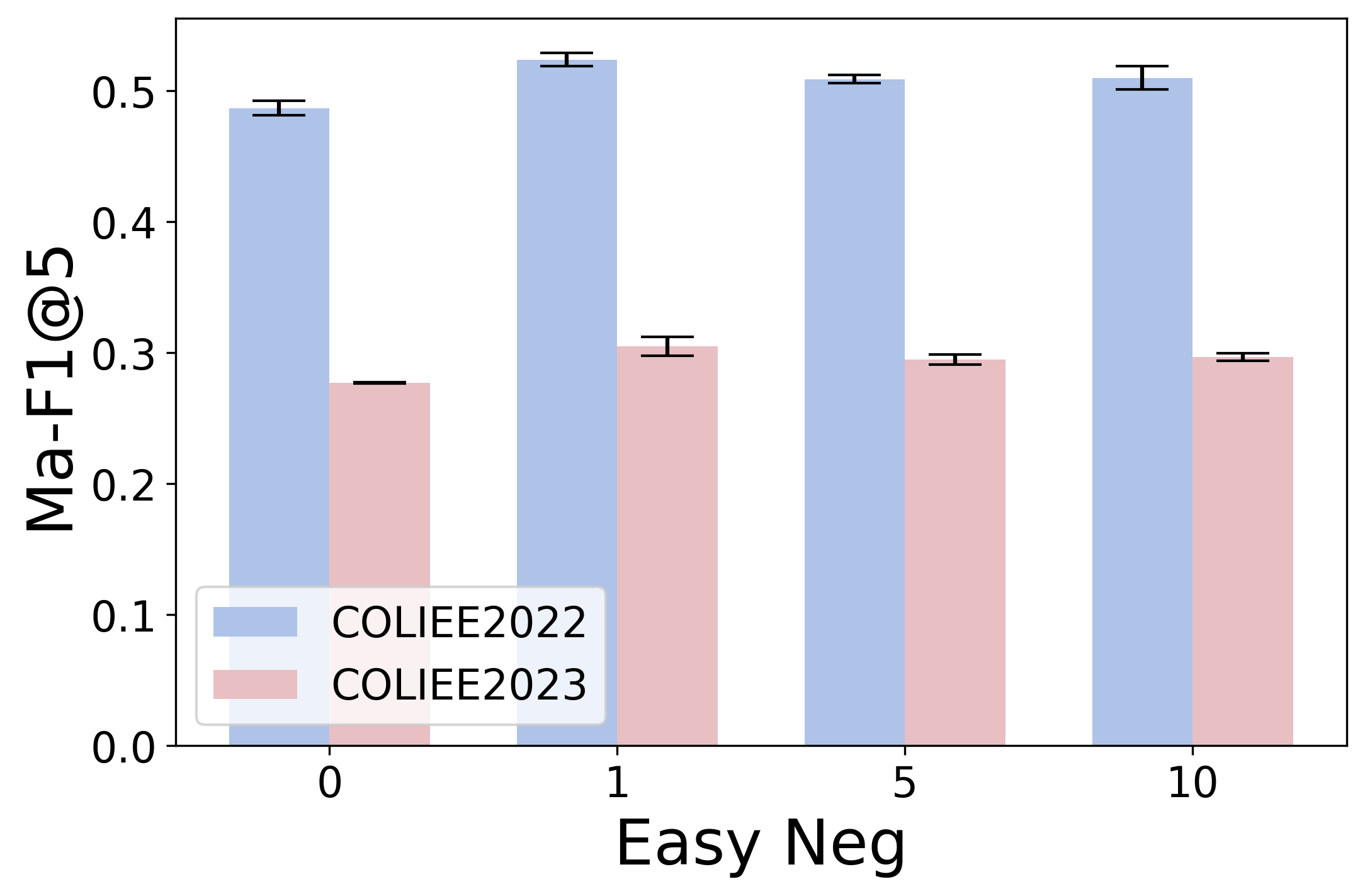}
    \label{fig:eneg_maF1}
}
\hfill
\subfigure[]{
    \includegraphics[width=0.45\linewidth]{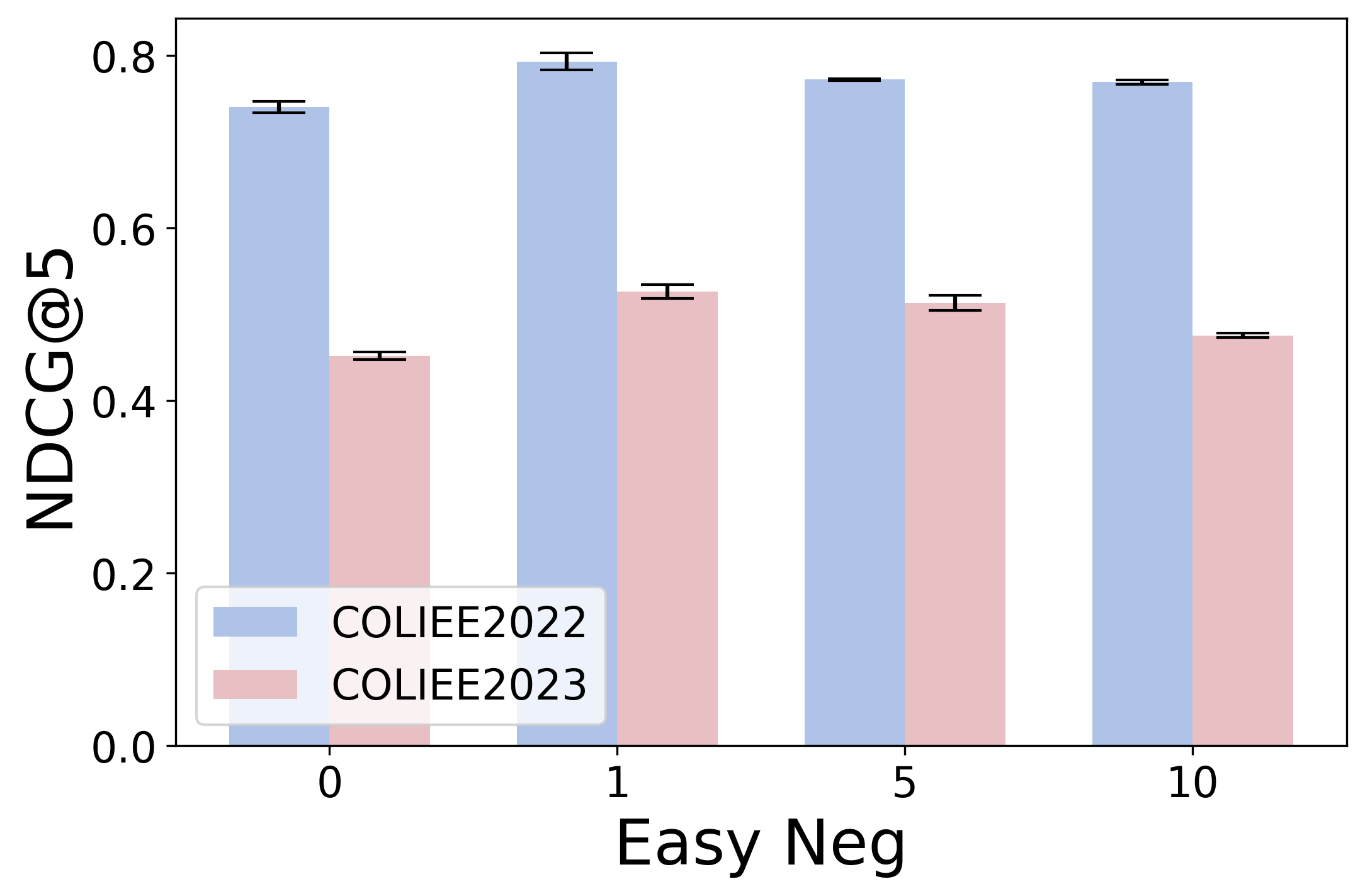}
    \label{fig:eneg_ndcg}
}
\caption{Parameter sensitivity for the number of easy negative samples in the contrastive loss of LEXA.}
\label{fig:easyneg}
\end{figure}

In this experiment, the temperature coefficient $\tau$ and the number of easy negative samples in the contrastive loss in Equation~(\ref{eq:cl}) are investigated for their parameter sensitivity in LEXA. Specifically, the result of the temperature coefficient $\tau$ is presented in Fig.~\ref{fig:temp}, and the results of the number of easy negative samples are presented in Fig.~\ref{fig:easyneg}.

\subsubsection{Temperature coefficient.} 
As illustrated in Fig.~\ref{fig:temp}, we evaluate the temperature parameter $\tau$ over the range of \{0.05,0.1,0.5,1,3\}. Across both COLIEE2022 and COLIEE2023, $\tau$=0.1 consistently yields the best Macro-F1 and NDCG results, indicating that a moderate temperature produces the most informative similarity distribution for contrastive learning. Larger temperatures (such as $\tau$=1 or $\tau$=3) reduce performance by overly smoothing the similarity scores, thereby diminishing the discriminative strength of the contrastive objective. In contrast, very small temperatures of $\tau$=0.05 make the similarity distribution too sharp, causing the model to over emphasize a few highly similar pairs and neglect useful relational structure among other samples. 

The results above demonstrate that the temperature parameter critically regulates the balance between smoothness and sharpness in the similarity space, with $\tau = 0.1$ offering the most stable and effective configuration for contrastive learning in LEXA.

\subsubsection{Number of easy negative samples} 
As shown in Fig.~\ref{fig:easyneg}, we investigate the effect of varying the number of easy negative samples using the set \{0, 1, 5, 10\}. Easy negatives are randomly selected non-relevant cases that help shape the contrastive learning signal. Across both datasets, introducing a small number of one easy negative sample leads to the best overall performance on Ma-F1@5 and NDCG@5, indicating that a limited amount of easy negative information strengthens the contrastive objective without overwhelming the model.
When no easy negatives are included, the performance on both datasets is clearly weaker, suggesting that the model benefits from the additional contrast provided by even a minimal set of easy negatives. However, using too many easy negatives (such as 5 or 10) results in a gradual decline in performance. This occurs because the learning objective becomes dominated by trivial negative instances, reducing the relative importance of informative positive and negative distinctions and weakening the supervision signal.

Overall, the results demonstrate that easy negatives are beneficial when used in moderation, and that a small number of such samples provides the most effective balance for training LEXA under the contrastive learning framework.

\section{Conclusion}
This paper introduced LEXA, an enhanced legal case retrieval framework that extends CaseGNN by fully utilising structural information in text-attributed case graphs. LEXA integrates three key improvements: an edge-updated graph attention layer (EUGAT) for jointly refining node and edge features, an enhanced graph contrastive learning objective for stronger supervision, and LLM-based contextualised embeddings to enrich graph representations. Experiments on COLIEE 2022 and COLIEE 2023 show that LEXA consistently outperforms CaseGNN and achieves state-of-the-art results. These findings highlight the importance of combining structural legal knowledge with contextual LLM representations to advance legal case retrieval.

\section*{Declarations}

\textbf{Funding declaration} This work is supported by Australian Research Council CE200100025, DP230101196,
DP230101753 and DE250100919.

\indent

\noindent\textbf{Ethics declaration} Not applicable.

\bibliographystyle{sn-mathphys-num}
\bibliography{sn-bibliography}

@inproceedings{LeCaRD,
  author       = {Yixiao Ma and
                  Yunqiu Shao and
                  Yueyue Wu and
                  Yiqun Liu and
                  Ruizhe Zhang and
                  Min Zhang and
                  Shaoping Ma},
  title        = {LeCaRD: {A} Legal Case Retrieval Dataset for Chinese Law System},
  booktitle    = {SIGIR},
  year         = {2021},
}

@inproceedings{COLIEE2023,
  author       = {Randy Goebel and Yoshinobu Kano and Mi-Young Kim and Maria Navas Loro and Nguyen Le Minh and Juliano Rabelo and Julien Rossi and Ken Satoh and Jaromir Savelka and Yunqiu Shao and Akira Shimazu and Satoshi Tojo and Vu Tran and Josef Valvoda and Hannes Westermann and Hiroaki Yamada and Masaharu Yoshioka and Sabine Wehnert},
  title        = {Competition on Legal Information Extraction/Entailment ({COLIEE})},
  year         = {2023},
}

@article{TF-IDF,
  author       = {Karen Sp{\"{a}}rck Jones},
  title        = {A statistical interpretation of term specificity and its application
                  in retrieval},
  journal      = {J. Documentation},
  volume       = {60},
  number       = {5},
  pages        = {493--502},
  year         = {2004},
}

@inproceedings{BM25,
  author       = {Stephen E. Robertson and
                  Steve Walker},
  title        = {Some Simple Effective Approximations to the 2-Poisson Model for Probabilistic
                  Weighted Retrieval},
  booktitle    = {{SIGIR}},
  year         = {1994},
}

@article{LMIR,
  author       = {Jay M. Ponte and
                  W. Bruce Croft},
  title        = {A Language Modeling Approach to Information Retrieval},
  journal      = {{SIGIR}},
  year         = {2017},
}

@inproceedings{BERT,
  author       = {Jacob Devlin and
                  Ming{-}Wei Chang and
                  Kenton Lee and
                  Kristina Toutanova},
  title        = {{BERT:} Pre-training of Deep Bidirectional Transformers for Language
                  Understanding},
  booktitle    = {{NAACL-HLT}},
  year         = {2019},
}

@article{RoBERTa,
  author       = {Yinhan Liu and
                  Myle Ott and
                  Naman Goyal and
                  Jingfei Du and
                  Mandar Joshi and
                  Danqi Chen and
                  Omer Levy and
                  Mike Lewis and
                  Luke Zettlemoyer and
                  Veselin Stoyanov},
  title        = {RoBERTa: {A} Robustly Optimized {BERT} Pretraining Approach},
  journal      = {CoRR},
  volume       = {abs/1907.11692},
  year         = {2019},
}

@article{DeepCT,
  author       = {Zhuyun Dai and
                  Jamie Callan},
  title        = {Context-Aware Sentence/Passage Term Importance Estimation For First
                  Stage Retrieval},
  journal      = {CoRR},
  volume       = {abs/1910.10687},
  year         = {2019},
}

@article{Law2Vec,
  author       = {Ilias Chalkidis and
                  Dimitrios Kampas},
  title        = {Deep learning in law: early adaptation and legal word embeddings trained
                  on large corpora},
  journal      = {Artif. Intell. Law},
  volume       = {27},
  number       = {2},
  pages        = {171--198},
  year         = {2019},
}

@article{Lawformer,
  author       = {Chaojun Xiao and
                  Xueyu Hu and
                  Zhiyuan Liu and
                  Cunchao Tu and
                  Maosong Sun},
  title        = {Lawformer: {A} pre-trained language model for Chinese legal long documents},
  journal      = {{AI} Open},
  volume       = {2},
  pages        = {79--84},
  year         = {2021},
}

@inproceedings{BERT-PLI,
  author    = {Yunqiu Shao and
               Jiaxin Mao and
               Yiqun Liu and
               Weizhi Ma and
               Ken Satoh and
               Min Zhang and
               Shaoping Ma},
  title     = {{BERT-PLI:} Modeling Paragraph-Level Interactions for Legal Case Retrieval},
  booktitle = {IJCAI},
  year      = {2020},
}

@inproceedings{MTFT-BERT,
  author       = {Amin Abolghasemi and
                  Suzan Verberne and
                  Leif Azzopardi},
  title        = {Improving BERT-based Query-by-Document Retrieval with Multi-task Optimization},
  booktitle    = {{ECIR}},
  year         = {2022},
}

@inproceedings{MVCL,
  author       = {Zhaowei Wang},
  title        = {Legal Element-oriented Modeling with Multi-view Contrastive Learning
                  for Legal Case Retrieval},
  booktitle    = {{IJCNN} },
  year         = {2022},
}

@article{LEGAL-BERT,
  author       = {Ilias Chalkidis and
                  Manos Fergadiotis and
                  Prodromos Malakasiotis and
                  Nikolaos Aletras and
                  Ion Androutsopoulos},
  title        = {{LEGAL-BERT:} The Muppets straight out of Law School},
  journal      = {CoRR},
  volume       = {abs/2010.02559},
  year         = {2020},
}

@article{SAILER,
  author       = {Haitao Li and
                  Qingyao Ai and
                  Jia Chen and
                  Qian Dong and
                  Yueyue Wu and
                  Yiqun Liu and
                  Chong Chen and
                  Qi Tian},
  title        = {{SAILER:} Structure-aware Pre-trained Language Model for Legal Case
                  Retrieval},
  journal      = {CoRR},
  volume       = {abs/2304.11370},
  year         = {2023},
}

@inproceedings{JOTR,
  author       = {Yixiao Ma and
                  Qingyao Ai and
                  Yueyue Wu and
                  Yunqiu Shao and
                  Yiqun Liu and
                  Min Zhang and
                  Shaoping Ma},
  title        = {Incorporating Retrieval Information into the Truncation of Ranking Lists for Better Legal Search},
  booktitle    = {{SIGIR}},
  year         = {2022},
}

@article{DoSSIER,
  author       = {Sophia Althammer and
                  Arian Askari and
                  Suzan Verberne and
                  Allan Hanbury},
  title        = {DoSSIER@COLIEE 2021: Leveraging dense retrieval and summarization-based
                  re-ranking for case law retrieval},
  journal      = {CoRR},
  volume       = {abs/2108.03937},
  year         = {2021},
}

@inproceedings{LEDsummary,
  author       = {Arian Askari and
                  Suzan Verberne},
  title        = {Combining Lexical and Neural Retrieval with Longformer-based Summarization
                  for Effective Case Law Retrieval},
  booktitle    = {DESIRES},
  series       = {{CEUR}},
  year         = {2021},
}

@inproceedings{BM25injtct,
  author       = {Arian Askari and
                  Amin Abolghasemi and
                  Gabriella Pasi and
                  Wessel Kraaij and
                  Suzan Verberne},
  title        = {Injecting the {BM25} Score as Text Improves BERT-Based Re-rankers},
  booktitle    = {{ECIR}},
  year         = {2023},
}

@article{LeiBi,
  author       = {Arian Askari and
                  Georgios Peikos and
                  Gabriella Pasi and
                  Suzan Verberne},
  title        = {LeiBi@COLIEE 2022: Aggregating Tuned Lexical Models with a Cluster-driven
                  BERT-based Model for Case Law Retrieval},
  journal      = {CoRR},
  volume       = {abs/2205.13351},
  year         = {2022},
}

@article{RPRS,
  author       = {Arian Askari and
                  Suzan Verberne and
                  Amin Abolghasemi and
                  Wessel Kraaij and
                  Gabriella Pasi},
  title        = {Retrieval for Extremely Long Queries and Documents with {RPRS:} a Highly Efficient and Effective Transformer-based Re-Ranker},
  journal      = {CoRR},
  volume       = {abs/2303.01200},
  year         = {2023},
}

@inproceedings{UA@COLIEE2022,
  author       = {Juliano Rabelo and
                  Mi{-}Young Kim and
                  Randy Goebel},
  title        = {Semantic-Based Classification of Relevant Case Law},
  booktitle    = {{JURISIN}},
  year         = {2022},
}

@article{NOWJ,
  author       = {Thi{-}Hai{-}Yen Vuong and
                  Hai{-}Long Nguyen and
                  Tan{-}Minh Nguyen and
                  Hoang{-}Trung Nguyen and
                  Thai{-}Binh Nguyen and
                  Ha{-}Thanh Nguyen},
  title        = {{NOWJ} at {COLIEE} 2023 - Multi-Task and Ensemble Approaches in Legal Information Processing},
  journal      = {CoRR},
  volume       = {abs/2306.04903},
  year         = {2023},
}

@inproceedings{LEVEN,
  author       = {Feng Yao and
                  Chaojun Xiao and
                  Xiaozhi Wang and
                  Zhiyuan Liu and
                  Lei Hou and
                  Cunchao Tu and
                  Juanzi Li and
                  Yun Liu and
                  Weixing Shen and
                  Maosong Sun},
  title        = {{LEVEN:} {A} Large-Scale Chinese Legal Event Detection Dataset},
  booktitle    = {{ACL}},
  year         = {2022},
}

@inproceedings{ConversationalAgent,
  author       = {Bulou Liu and
                  Yiran Hu and
                  Yueyue Wu and
                  Yiqun Liu and
                  Fan Zhang and
                  Chenliang Li and
                  Min Zhang and
                  Shaoping Ma and
                  Weixing Shen},
  title        = {Investigating Conversational Agent Action in Legal Case Retrieval},
  booktitle    = {{ECIR}},
  year         = {2023},
}

@inproceedings{JNLP@COLIEE2019,
  author       = {Vu D. Tran and
                  Minh Le Nguyen and
                  Ken Satoh},
  title        = {Building Legal Case Retrieval Systems with Lexical Matching and Summarization
                  using {A} Pre-Trained Phrase Scoring Model},
  booktitle    = {{ICAIL}},
  year         = {2019},
}

@article{CL4LJP,
author = {Zhang, Han and Dou, Zhicheng and Zhu, Yutao and Wen, Ji-Rong},
title = {Contrastive Learning for Legal Judgment Prediction},
year = {2023},
volume = {41},
number = {4},
journal = {ACM Trans. Inf. Syst.},
pages = {25},
}

@inproceedings{QAjudge,
  author       = {Haoxi Zhong and
                  Yuzhong Wang and
                  Cunchao Tu and
                  Tianyang Zhang and
                  Zhiyuan Liu and
                  Maosong Sun},
  title        = {Iteratively Questioning and Answering for Interpretable Legal Judgment
                  Prediction},
  booktitle    = {{AAAI}},
  year         = {2020},
}

@inproceedings{IOT-Match,
  author       = {Weijie Yu and
                  Zhongxiang Sun and
                  Jun Xu and
                  Zhenhua Dong and
                  Xu Chen and
                  Hongteng Xu and
                  Ji{-}Rong Wen},
  title        = {Explainable Legal Case Matching via Inverse Optimal Transport-based
                  Rationale Extraction},
  booktitle    = {{SIGIR}},
  year         = {2022},
}

@article{Law-Match,
  author       = {Zhongxiang Sun and
                  Jun Xu and
                  Xiao Zhang and
                  Zhenhua Dong and
                  Ji{-}Rong Wen},
  title        = {Law Article-Enhanced Legal Case Matching: a Model-Agnostic Causal
                  Learning Approach},
  journal      = {CoRR},
  volume       = {abs/2210.11012},
  year         = {2022},
}

@article{longformer,
  author       = {Iz Beltagy and
                  Matthew E. Peters and
                  Arman Cohan},
  title        = {Longformer: The Long-Document Transformer},
  journal      = {CoRR},
  volume       = {abs/2004.05150},
  year         = {2020},
}

@inproceedings{monot5,
    title = "Document Ranking with a Pretrained Sequence-to-Sequence Model",
    author = "Nogueira, Rodrigo  and
      Jiang, Zhiying  and
      Pradeep, Ronak  and
      Lin, Jimmy",
    booktitle = {EMNLP},
    year = "2020",
}

@article{query_conversational_agent,
  author       = {Bulou Liu and
                  Yueyue Wu and
                  Fan Zhang and
                  Yiqun Liu and
                  Zhihong Wang and
                  Chenliang Li and
                  Min Zhang and
                  Shaoping Ma},
  title        = {Query Generation and Buffer Mechanism: Towards a better conversational
                  agent for legal case retrieval},
  journal      = {Inf. Process. Manag.},
  year         = {2022},
}

@article{T5,
  author       = {Colin Raffel and
                  Noam Shazeer and
                  Adam Roberts and
                  Katherine Lee and
                  Sharan Narang and
                  Michael Matena and
                  Yanqi Zhou and
                  Wei Li and
                  Peter J. Liu},
  title        = {Exploring the Limits of Transfer Learning with a Unified Text-to-Text
                  Transformer},
  journal      = {J. Mach. Learn. Res.},
  year         = {2020},
}

@inproceedings{promptcase,
  title={Prompt-based effective input reformulation for legal case retrieval},
  author={Tang, Yanran and Qiu, Ruihong and Li, Xue},
  booktitle={Australasian database conference},
  year={2023},
}

@inproceedings{transformer,
  author       = {Ashish Vaswani and
                  Noam Shazeer and
                  Niki Parmar and
                  Jakob Uszkoreit and
                  Llion Jones and
                  Aidan N. Gomez and
                  Lukasz Kaiser and
                  Illia Polosukhin},
  title        = {Attention is All you Need},
  booktitle    = {NeurIPS},
  pages        = {5998--6008},
  year         = {2017},
}

@inproceedings{COLIEE2022,
  author       = {Randy Goebel and Yoshinobu Kano and Mi-Young Kim and Maria Navas Loro and Nguyen Le Minh and Juliano Rabelo and Julien Rossi and Ken Satoh and Jaromir Savelka and Yunqiu Shao and Akira Shimazu and Satoshi Tojo and Vu Tran and Josef Valvoda and Hannes Westermann and Hiroaki Yamada and Masaharu Yoshioka and Sabine Wehnert},
  title        = {Competition on Legal Information Extraction/Entailment ({COLIEE})},
  year         = {2022},
}

@inproceedings{GCN,
  author       = {Thomas N. Kipf and
                  Max Welling},
  title        = {Semi-Supervised Classification with Graph Convolutional Networks},
  booktitle    = {{ICLR}},
  year         = {2017},
}

@inproceedings{GAT,
  author       = {Petar Velickovic and
                  Guillem Cucurull and
                  Arantxa Casanova and
                  Adriana Romero and
                  Pietro Li{\`{o}} and
                  Yoshua Bengio},
  title        = {Graph Attention Networks},
  booktitle    = {{ICLR}},
  year         = {2018},
}

@inproceedings{OpenIE,
  author       = {Gabor Angeli and
                  Melvin Jose Johnson Premkumar and
                  Christopher D. Manning},
  title        = {Leveraging Linguistic Structure For Open Domain Information Extraction},
  booktitle    = {{ACL}},
  year         = {2015},
}

@article{dropout,
  author       = {Nitish Srivastava and
                  Geoffrey E. Hinton and
                  Alex Krizhevsky and
                  Ilya Sutskever and
                  Ruslan Salakhutdinov},
  title        = {Dropout: a simple way to prevent neural networks from overfitting},
  journal      = {J. Mach. Learn. Res.},
  year         = {2014},
}

@inproceedings{Adam,
  author       = {Diederik P. Kingma and
                  Jimmy Ba},
  title        = {Adam: {A} Method for Stochastic Optimization},
  booktitle    = {ICLR},
  year         = {2015},
}

@article{puma,
  title={Puma: Efficient continual graph learning for node classification with graph condensation},
  author={Liu, Yilun and Qiu, Ruihong and Tang, Yanran and Yin, Hongzhi and Huang, Zi},
  journal={IEEE Transactions on Knowledge and Data Engineering},
  volume={37},
  number={1},
  pages={449--461},
  year={2024},
  publisher={IEEE}
}

@inproceedings{casegnn,
  author       = {Yanran Tang and
                  Ruihong Qiu and
                  Yilun Liu and
                  Xue Li and
                  Zi Huang},
  title        = {CaseGNN: Graph Neural Networks for Legal Case Retrieval with Text-Attributed Graphs},
  booktitle    = {ECIR},
  year         = {2024},
}

@inproceedings{moco,
  author       = {Kaiming He and
                  Haoqi Fan and
                  Yuxin Wu and
                  Saining Xie and
                  Ross B. Girshick},
  title        = {Momentum Contrast for Unsupervised Visual Representation Learning},
  booktitle    = {CVPR},
  pages        = {9726--9735},
  year         = {2020},
}

@inproceedings{SimCLR,
  author       = {Ting Chen and
                  Simon Kornblith and
                  Mohammad Norouzi and
                  Geoffrey E. Hinton},
  title        = {A Simple Framework for Contrastive Learning of Visual Representations},
  booktitle    = {ICML},
  pages        = {1597--1607},
  year         = {2020},
}

@inproceedings{NCE,
  author       = {Michael Gutmann and
                  Aapo Hyv{\"{a}}rinen},
  title        = {Noise-contrastive estimation: {A} new estimation principle for unnormalized
                  statistical models},
  booktitle    = {AISTATS},
  pages        = {297--304},
  year         = {2010},
}

@inproceedings{NCE4NPLMs,
  author       = {Andriy Mnih and
                  Yee Whye Teh},
  title        = {A fast and simple algorithm for training neural probabilistic language
                  models},
  booktitle    = {ICML},
  year         = {2012}
}

@inproceedings{SimCSE,
  author       = {Tianyu Gao and
                  Xingcheng Yao and
                  Danqi Chen},
  editor       = {Marie{-}Francine Moens and
                  Xuanjing Huang and
                  Lucia Specia and
                  Scott Wen{-}tau Yih},
  title        = {SimCSE: Simple Contrastive Learning of Sentence Embeddings},
  booktitle    = {EMNLP},
  pages        = {6894--6910},
  year         = {2021}
}

@article{CLEAR,
  author       = {Zhuofeng Wu and
                  Sinong Wang and
                  Jiatao Gu and
                  Madian Khabsa and
                  Fei Sun and
                  Hao Ma},
  title        = {{CLEAR:} Contrastive Learning for Sentence Representation},
  journal      = {CoRR},
  volume       = {abs/2012.15466},
  year         = {2020}
}

@inproceedings{COCO-LM,
  author       = {Yu Meng and
                  Chenyan Xiong and
                  Payal Bajaj and
                  Saurabh Tiwary and
                  Paul Bennett and
                  Jiawei Han and
                  Xia Song},
  title        = {{COCO-LM:} Correcting and Contrasting Text Sequences for Language
                  Model Pretraining},
  booktitle    = {NeurIPS},
  pages        = {23102--23114},
  year         = {2021}
}

@inproceedings{PSSL,
  author       = {Yujia Zhou and
                  Zhicheng Dou and
                  Yutao Zhu and
                  Ji{-}Rong Wen},
  title        = {{PSSL:} Self-supervised Learning for Personalized Search with Contrastive
                  Sampling},
  booktitle    = {CIKM},
  pages        = {2749--2758},
  year         = {2021}
}

@inproceedings{ANCE,
  author       = {Lee Xiong and
                  Chenyan Xiong and
                  Ye Li and
                  Kwok{-}Fung Tang and
                  Jialin Liu and
                  Paul N. Bennett and
                  Junaid Ahmed and
                  Arnold Overwijk},
  title        = {Approximate Nearest Neighbor Negative Contrastive Learning for Dense
                  Text Retrieval},
  booktitle    = {ICLR},
  year         = {2021}
}

@inproceedings{DropEdge,
  author       = {Yu Rong and
                  Wenbing Huang and
                  Tingyang Xu and
                  Junzhou Huang},
  title        = {DropEdge: Towards Deep Graph Convolutional Networks on Node Classification},
  booktitle    = {ICLR},
  year         = {2020}
}

@inproceedings{GRAND,
  author       = {Wenzheng Feng and
                  Jie Zhang and
                  Yuxiao Dong and
                  Yu Han and
                  Huanbo Luan and
                  Qian Xu and
                  Qiang Yang and
                  Evgeny Kharlamov and
                  Jie Tang},
  title        = {Graph Random Neural Networks for Semi-Supervised Learning on Graphs},
  booktitle    = {NeurIPS},
  year         = {2020}
}

@inproceedings{GraphCL,
  author       = {Yuning You and
                  Tianlong Chen and
                  Yongduo Sui and
                  Ting Chen and
                  Zhangyang Wang and
                  Yang Shen},
  title        = {Graph Contrastive Learning with Augmentations},
  booktitle    = {NeurIPS},
  year         = {2020}
}

@inproceedings{BGRL,
  author       = {Shantanu Thakoor and
                  Corentin Tallec and
                  Mohammad Gheshlaghi Azar and
                  Mehdi Azabou and
                  Eva L. Dyer and
                  R{\'{e}}mi Munos and
                  Petar Velickovic and
                  Michal Valko},
  title        = {Large-Scale Representation Learning on Graphs via Bootstrapping},
  booktitle    = {ICLR},
  year         = {2022}
}

@article{GraphCrop,
  author       = {Yiwei Wang and
                  Wei Wang and
                  Yuxuan Liang and
                  Yujun Cai and
                  Bryan Hooi},
  title        = {GraphCrop: Subgraph Cropping for Graph Classification},
  journal      = {CoRR},
  volume       = {abs/2009.10564},
  year         = {2020}
}

@inproceedings{word2vec,
  author       = {Tom{\'{a}}s Mikolov and
                  Ilya Sutskever and
                  Kai Chen and
                  Gregory S. Corrado and
                  Jeffrey Dean},
  title        = {Distributed Representations of Words and Phrases and their Compositionality},
  booktitle    = {NeurIPS},
  year         = {2013},
}

@inproceedings{caselink,
  title={Caselink: Inductive graph learning for legal case retrieval},
  author={Tang, Yanran and Qiu, Ruihong and Yin, Hongzhi and Li, Xue and Huang, Zi},
  booktitle={SIGIR},
  year={2024}
}

@article{mmteb,
  author = {Kenneth Enevoldsen and Isaac Chung and Imene Kerboua and Márton Kardos and Ashwin Mathur and David Stap and Jay Gala and Wissam Siblini and Dominik Krzemiński and Genta Indra Winata and Saba Sturua and Saiteja Utpala and Mathieu Ciancone and Marion Schaeffer and Gabriel Sequeira and Diganta Misra and Shreeya Dhakal and Jonathan Rystrøm and Roman Solomatin and Ömer Çağatan and Akash Kundu and Martin Bernstorff and Shitao Xiao and Akshita Sukhlecha and Bhavish Pahwa and Rafał Poświata and Kranthi Kiran GV and Shawon Ashraf and Daniel Auras and Björn Plüster and Jan Philipp Harries and Loïc Magne and Isabelle Mohr and Mariya Hendriksen and Dawei Zhu and Hippolyte Gisserot-Boukhlef and Tom Aarsen and Jan Kostkan and Konrad Wojtasik and Taemin Lee and Marek Šuppa and Crystina Zhang and Roberta Rocca and Mohammed Hamdy and Andrianos Michail and John Yang and Manuel Faysse and Aleksei Vatolin and Nandan Thakur and Manan Dey and Dipam Vasani and Pranjal Chitale and Simone Tedeschi and Nguyen Tai and Artem Snegirev and Michael Günther and Mengzhou Xia and Weijia Shi and Xing Han Lù and Jordan Clive and Gayatri Krishnakumar and Anna Maksimova and Silvan Wehrli and Maria Tikhonova and Henil Panchal and Aleksandr Abramov and Malte Ostendorff and Zheng Liu and Simon Clematide and Lester James Miranda and Alena Fenogenova and Guangyu Song and Ruqiya Bin Safi and Wen-Ding Li and Alessia Borghini and Federico Cassano and Hongjin Su and Jimmy Lin and Howard Yen and Lasse Hansen and Sara Hooker and Chenghao Xiao and Vaibhav Adlakha and Orion Weller and Siva Reddy and Niklas Muennighoff},
  journal      = {CoRR},
  volume       = {abs/2502.13595},
  title = {MMTEB: Massive Multilingual Text Embedding Benchmark},
  year = {2025},
}

@article{e5mistral,
  title={Text Embeddings by Weakly-Supervised Contrastive Pre-training},
  author={Wang, Liang and Yang, Nan and Huang, Xiaolong and Jiao, Binxing and Yang, Linjun and Jiang, Daxin and Majumder, Rangan and Wei, Furu},
  journal      = {CoRR},
  volume       = {abs/2212.03533},
  year={2022}
}

@article{qwen3embedding,
  title={Qwen3 Embedding: Advancing Text Embedding and Reranking Through Foundation Models},
  author={Zhang, Yanzhao and Li, Mingxin and Long, Dingkun and Zhang, Xin and Lin, Huan and Yang, Baosong and Xie, Pengjun and Yang, An and Liu, Dayiheng and Lin, Junyang and Huang, Fei and Zhou, Jingren},
  journal      = {CoRR},
  volume       = {abs/2506.05176},
  year={2025}
}

@inproceedings{GIN,
  author       = {Keyulu Xu and
                  Weihua Hu and
                  Jure Leskovec and
                  Stefanie Jegelka},
  title        = {How Powerful are Graph Neural Networks?},
  booktitle    = {ICLR},
  year         = {2019},
}

@misc {inf-retriever-v1,
    author       = {Junhan Yang and Jiahe Wan and Yichen Yao and Wei Chu and Yinghui Xu and Yuan Qi},
    title        = { inf-retriever-v1 (Revision 5f469d7) },
    year         = {2025},
    url          = { https://huggingface.co/infly/inf-retriever-v1 },
    publisher    = { Hugging Face },
}

@inproceedings{uqlegalai,
  title={UQLegalAI@ COLIEE2025: advancing legal case retrieval with large language models and graph neural networks},
  author={Tang, Yanran and Qiu, Ruihong and Huang, Zi},
  booktitle={COLIEE},
  year={2025}
}

@inproceedings{reakase,
  title={ReaKase-8B: Legal Case Retrieval via Knowledge and Reasoning Representations with LLMs},
  author={Tang, Yanran and Qiu, Ruihong and Li, Xue and Huang, Zi},
  booktitle={Australasian Database Conference},
  year={2025},
}

@inproceedings{tnt-ood,  author       = {Danny Wang and
                  Ruihong Qiu and
                  Guangdong Bai and
                  Zi Huang},
  title        = {Text Meets Topology: Rethinking Out-of-distribution Detection in Text-Rich Networks},
  booktitle    = {EMNLP},
  year         = {2025},
}

\end{document}